\documentclass[11pt, a4paper]{article}   	

\usepackage{geometry}                		
\usepackage{graphicx}				

\usepackage{hyperref}
\usepackage{amssymb}
\usepackage{tikz}
\usepackage{latexsym, amsmath, mathrsfs}
\usepackage[utf8]{inputenc}
\usepackage{authblk}
\usepackage{xcolor}
\usepackage{esint}
\usepackage{empheq}
\usepackage{subcaption}
\usepackage[normalem]{ulem}

\usepackage{cite}
\usepackage{hyperref}

\newcommand{\atan}{\textrm{atan}}

\newcommand{\oneph}{1{\rm ph}}
\newcommand{\twoph}{2{\rm ph}}
\newcommand{\threeph}{3{\rm ph}}

\newcommand{\bfp}{\mathbf{p}}
\newcommand{\bfh}{\mathbf{h}}
\newcommand{\bfa}{\boldsymbol{\alpha}}

\newcommand{\Tdr}{T^{\rm dr}}

\begin{document}

\title{The two particle-hole pairs contribution to the dynamic correlation functions of quantum integrable models}

\author{Miłosz Panfil\\
\small{Faculty of Physics, University of Warsaw, ul. Pasteura 5, 02-093 Warsaw, Poland}
}

\date{\today}	

\maketitle						



\begin{abstract}
We consider the problem of computing dynamic correlation functions of quantum integrable models employing the thermodynamic form-factor approach. Specifically, we focus on correlations of local operators that conserve the number of particles and consider the 2-particle-hole contribution to their two-point functions. With the method developed being generally applicable to any finite energy and entropy state, our primary focus  is on the thermal states. To exemplify this approach, we chose the Lieb-Liniger model and study the leading contribution from 2-particle-hole excitations at small momenta to the dynamic density-density correlation function. We also consider analogous contributions to two-point functions of higher local conserved densities and currents present in integrable theories.
\end{abstract}

\section{Introduction}

In this work, we consider the problem of computing dynamic correlation functions in quantum integrable models. Our focus is on thermodynamically large systems at finite energy and entropy density; a typical example is the state of thermal equilibrium at a non-zero temperature. Recent years have witnessed various developments in this direction. These include the microscopic approach based on the ABACUS method~\cite{1742-5468-2007-01-P01008,2006_Caux_PRA_74,2014_PRA_Panfil}; the computation of long-distance and large-time asymptotics in and out of the equilibrium in quantum many-body systems~\cite{Kozlowski_2015,G_hmann_2017,2018ScPP....5...54D,Doyon_drude,G_hmann_2020_1,G_hmann_2020_2};
the computation of correlation functions in the setup of Integrable Quantum Field Theories at a finite temperature~\cite{LECLAIR1999624,CastroAlvaredo:2002ud,Doyon:2005jf,2007SIGMA...3..011D,Essler_2009,1742-5468-2010-11-P11012} and out of the equilibrium~\cite{2010_Fioretto_NJP_12,Bertini_2014,Cort_s_Cubero_2017,2018PhRvL.121k0402K,2020ScPP....9...33G}. As far as the latter is concerned, an interesting result was the derivation of a closed-form expression for one-point functions in an arbitrary state of the system~\cite{Negro_2013,Bastianello_2018_1,Bastianello_2018_2}. 

In this work, we approach the correlation functions using the thermodynamic form-factors~\cite{Smooth_us,SciPostPhys.1.2.015,2018JSMTE..03.3102D,Bootstrap_JHEP,Cortes_Cubero_2020} and focus on the Lieb-Liniger model. Apart from a general interest in the computation of correlation functions in strongly correlated quantum systems, the finite temperature dynamic correlation functions in the Lieb-Liniger model are important, e.g. for Bragg spectroscopy in cold atomic gases~\cite{2001_Brunello_PRA_64,2011_Fabbri_PRA_83,PhysRevA.91.043617,PhysRevLett.115.085301}. 

Another incentive to develop general methods of computing correlation functions comes from experimental progress making it possible to probe the nonequilibrium dynamics with cold atomic gases~\cite{Kinoshita:2006aa,PhysRevLett.96.136801,2008Natur.452..854R,2011RvMP...83..863P,Eisert:2015aa,doi:10.1146/annurev-conmatphys-031214-014548}. In that respect, the thermodynamic form-factor approach has already shown some utility, leading, for example, to the concept of generalized detailed balance~\cite{De_Nardis_2017,Foini_2017,SciPostPhys.1.2.015} and to the prediction of edge singularities in correlation functions of certain nonequilibrium steady states~\cite{De_Nardis_2018}.

Paralleling these various developments was the introduction of Generalized Hydrodynamics (GHD)~\cite{PhysRevX.6.041065,2016PhRvL.117t7201B,doyon2019lecture}, an effective theory for integrable models in inhomogeneous setups. It led to a burst of activities in formulating GHD for various models and setups, including the Lieb-Liniger gas~\cite{2017arXiv171100873C,2017PhRvL.119s5301D,2017PhRvL.119v0604B,2018ScPP....5...54D,Panfil_2019,Bastianello_2019,Franchini_2016, 2020arXiv200704861B} and related continuum models~\cite{2018PhRvL.120d5301D,Bastianello_2018} with its validity being recently confirmed experimentally~\cite{2019PhRvL.122i0601S}.  GHD was originally formulated only at the ballistic level, with the diffusive effects being included later~\cite{2018PhRvL.121p0603D,10.21468/SciPostPhys.6.4.049,Gopalakrishnan:2018wyg}, partially with an input provided by the thermodynamic form-factors.

Finally, the thermodynamic bootstrap program (TBP)~\cite{Bootstrap_JHEP,Cortes_Cubero_2020}, which generalizes the vacuum form-factors program~\cite{doi:10.1142/1115} of Integrable Quantum Field Theories to finite density states, hints at a universal structure behind thermodynamic form-factors. Its predictions, combined with the quench action approach~\cite{1742-5468-2016-6-064006,2013_Caux_PRL_110}, led to a reconstruction of the GHD~\cite{cubero2020generalized}.

To introduce the main aim of this work, we start with an outline of the thermodynamic form-factor approach applied to the repulsive Lieb-Liniger model.
This approach relies on the spectral representation and involves the particle picture of an integrable theory: the eigenstate of the system is characterized once its particle content is specified. Denoting $|\vartheta \rangle$ a state of the system with a particle content labeled by $\vartheta$ we are interested in the computation of the (connected) two-point function 
\begin{equation}
S(x,t) = \langle \vartheta| \mathcal{O}(x,t) \mathcal{O}(0)| \vartheta\rangle,
\end{equation}
of some Hermitian operator $\mathcal{O}(x)$, or its Fourier transform $S(k,\omega)$. For local operators conserving the number of particles, the correlation function, due to the presence of an effective Pauli principle, is a sum of contributions with a fixed number of particle-hole excitations,
\begin{equation}
	S(k, \omega) = \sum_{m=1}^{\infty} S^{m{\rm ph}}(k,\omega),
\end{equation}
where the $m$ particle-hole contribution ($m{\rm ph}$ in short) is
\begin{equation}
	S^{m{\rm ph}}(k,\omega) = \frac{(2\pi)^2}{(m!)^2}  \fint {\rm d}\bfp_m {\rm d} \bfh_m |\langle \vartheta| \mathcal{O}(0) | \vartheta, \bfp, \bfh \rangle|^2 \delta( \omega - \omega(\bfp, \bfh)) \delta( k - k(\bfp, \bfh)). \label{spectral_intro}
\end{equation}
The sum is performed over all possible states containing $m$ pairs of excitations created on top of $|\vartheta\rangle$. We denote the corresponding state by $|\vartheta, \bfp, \bfh\rangle$, with the notation $\bfp_m = \{p_j\}_{j=1}^m$ abbreviated to $\bfp$ when the cardinality of the set is clear from the context or irrelevant. The positions of particles $\bfp$ and holes $\bfh$ are parametrized by real numbers. The particles and holes carry momentum and energy. We denote $\omega(\bfp, \bfh)$ and $k(\bfp, \bfh)$ the total energy and momentum of the state $|\vartheta, \bfp, \bfh\rangle$ with respect to $|\vartheta\rangle$. Finally, $\langle \vartheta| \mathcal{O}(0) | \vartheta, \bfp, \bfh \rangle$ are thermodynamic form factors of the operator $\mathcal{O}(0)$.

The form-factors in~\eqref{spectral_intro}, with 2 or more particle-hole excitations, have simple poles, called annihilation or kinematic poles, whenever $h_i = p_j$. These singularities are regularized by adopting the Hadamard regularization for integrals over particles (holes) positions. The integrals over the holes (particles) positions are performed afterward and are regular. The Hadamard regularization assigns a finite value
\begin{equation}
	\fint_0 {\rm d}x f(x) = \lim_{\epsilon \rightarrow 0^+} \left(\int {\rm d}x f(x) \Theta(|x| - \epsilon) - \frac{2}{\epsilon} \lim_{x\rightarrow 0} \left(x^2 f(x) \right) \right),
\end{equation}
to a function $f(x)$ having a double pole (here assumed to be at $x=0$) in the region of the integration. We use the subscript below the integration symbol to denote the position of the pole when necessary. Unless otherwise stated, all the integrals are along the real line.

The applicability of this approach to the dynamic correlation function relies on the access to the ingredients of~\eqref{spectral_intro}. In the context of integrable models, the functions $\omega(\bfp, \bfh)$ and $k(\bfp, \bfh)$ follow from the standard construction of Thermodynamic Bethe Ansatz~\cite{1969_Yang_JMP_10,KorepinBOOK,TakahashiBOOK,2012_Mossel_JPA_45,2012PhRvL.109q5301C}. On the other hand, the thermodynamic form-factors are known only in a few cases. The notable examples are the thermodynamic form-factor of the density operator in the Lieb-Liniger model~\cite{Smooth_us,2018JSMTE..03.3102D}, the small particle-hole number form-factors in the small momentum limit of conserved charges and currents for a generic integrable model~\cite{Doyon_drude,2018ScPP....5...54D}, and form-factors of the vertex operators in the Sinh-Gordon Integrable Quantum Field Theory~\cite{Bootstrap_JHEP}. 

From this limited number of examples arose hints on the universal structure of the thermodynamic form-factors whenever particles and holes are close to each other. In the TBP, this universal structure is captured by the annihilation axiom~\cite{Bootstrap_JHEP} and states that form-factors involving two or more pairs of particle-hole excitations have simple poles. These singularities have a deep physical meaning, signalling a large contribution of those weakly excited particle-hole pairs to the spectral sum. On the other hand, when the particle and hole are exactly on top of each other, there is no particle-hole excitation and potential contributions to the correlation function should not be included. The solution to this problem comes from adopting a regularization scheme~\cite{Smooth_us,10.21468/SciPostPhys.6.4.049,Bootstrap_JHEP} which can be recast in the form of Hadamard integrals presented above. 

Given the universal pole structure of the form-factors, the question then arises of how to perform the spectral sum in practice. 
In this work, we address this problem by focusing on the simplest non-trivial case of $\twoph$ contributions.\footnote{The $\oneph$ form-factors do not have singularities when $p \rightarrow h$ and moreover the presence of the $\delta$-functions fixes uniquely $p$ and $h$ so there is no summation.}  Furthermore, we consider only the leading singular part of form-factors as this is the one that requires the Hadamard regularization and it constitutes the leading contribution to the correlation function  from $\twoph$ excitations at small momentum~$k$. To be specific, we consider the Lieb-Liniger model and density-density correlation functions (with a straightforward generalization to two-point functions of higher conserved charges and currents). However, the methods developed are applicable to other quantum integrable models including Integrable Quantum Field Theories.

In Section~\ref{sec:LL} we recall the ingredients of the solution to the Lieb-Liniger. In the following Section~\ref{sec:small}, we formalize the concept of small particle-hole excitations. In Section~\ref{sec:2ph} we present the main result, the computation of the $\twoph$ contribution for small excitations. In the following two sections, we apply this result in two contexts: first, to compute the same type of contribution to two-point functions of higher conserved densities and currents (Section~\ref{sec:higher}), second, to compute the density-density correlation function in the ground state (Section~\ref{sec:gs}). We finish with the conclusions. Appendices are devoted to complementary computations. In Appendix~\ref{app:1ph} we recall the computation of the $\oneph$ contribution to the density-density correlation function. In Appendix~\ref{app:3ph}, based on the TBP and $\twoph$ form-factors of the density in the Lieb-Liniger model, we conjecture an expression for $\threeph$ form-factors. In Appendix~\ref{app:Hadamard} we fill in a small gap in the derivation of the Hadamard regularization scheme for the spectral sum.

\section{Dynamic correlation functions and Lieb-Liniger gas} \label{sec:LL}

The Lieb-Liniger model is defined by the following Hamiltonian~\cite{1963_Lieb_PR_130_1,1963_Lieb_PR_130_2}
\begin{equation}
	H = - \sum_{j=1}^N \frac{\partial^2}{\partial x_j^2} + 2c \sum_{j>k}^N \delta(x_j - x_k),
\end{equation}
in units where $\hbar = 1$, $2m = 1$ and with $c$ being the interaction parameter. We consider only repulsive interactions ($c>0$). 
The standard Bethe Ansatz techniques provide solution to the system with a finite number of particles $N$~\cite{1963_Lieb_PR_130_1,1963_Lieb_PR_130_2,KorepinBOOK}. Here, we are concerned with the Thermodynamic Bethe Ansatz~\cite{1969_Yang_JMP_10,KorepinBOOK,TakahashiBOOK} solution valid in thermodynamic limit when $L \rightarrow \infty$ with the density of particles $N/L$ fixed. In that case, the eigenstates of the system are described by the filling function $\vartheta(\lambda)$ and related density of particles $\rho_p(\lambda)$, with $ L \rho_p(\lambda) {\rm d}\lambda$ the number of particles in range $[\lambda, \lambda + {\rm d}\lambda]$.\footnote{We note that the thermodynamic form-factors depend on the state of the system only through $\vartheta(\lambda)$ and therefore the filling function carries sufficient amount of information to characterize the state of the system for our purpose. However, as pointed out in~\cite{2020arXiv200715396G}, it is possible that in some contexts the filling function is not enough and more refined information is needed.} The density of particles is given by
\begin{equation}
	\rho_p(\lambda) = \vartheta(\lambda) \rho_t(\lambda),
\end{equation}
where $\rho_t(\lambda)$ is a solution to the linear integral equation
\begin{equation}
	\rho_t(\lambda) = \frac{1}{2\pi} + \int {\rm d}\lambda' \vartheta(\lambda')\, T(\lambda, \lambda') \rho_t(\lambda'),
\end{equation}
with the differential scattering kernel 
\begin{equation}
	T(\lambda, \lambda') = \frac{1}{\pi} \frac{c}{(\lambda - \lambda')^2 + c^2}.
\end{equation}
For the future convenience, we also define the density of holes 
\begin{equation}
	\rho_h(\lambda) = (1 - \vartheta(\lambda)) \rho_t(\lambda) = \rho_t(\lambda) - \rho_p(\lambda).
\end{equation}

The density of particles $\rho_p(\lambda)$ determines various macroscopic variables of the state like the total density
\begin{equation}
	\frac{N}{L} = \int {\rm d}\lambda\, \rho_p(\lambda),
\end{equation}
and its energy and momentum
\begin{equation}
	E[\vartheta] = L \int {\rm d}\lambda\, \rho_p(\lambda) \lambda^2, \qquad P[\vartheta] = L \int_{-\infty}^{\infty} {\rm d}\lambda\, \rho_p(\lambda) \lambda. 
\end{equation}
We assign to a filling function $\vartheta(\lambda)$ a state $|\vartheta\rangle$. It can be constructed from (normalized) sum of its microscopic realizations~\cite{Smooth_us,1742-5468-2016-6-064006,2018JSMTE..03.3102D,Bootstrap_JHEP}. Their number is counted by (the exponential of) the entropy~\cite{1969_Yang_JMP_10}
\begin{equation}
	S[\vartheta] = L \int {\rm d}\lambda \left( \rho_t(\lambda) \ln \rho_t(\lambda) - \rho_p(\lambda) \ln \rho_p(\lambda) - \rho_h(\lambda) \ln \rho_h(\lambda) \right).
\end{equation}
We are then interested in the dynamic correlation functions
\begin{equation}
	\langle \vartheta| \mathcal{O}(x,t) \mathcal{O}(0) |\vartheta \rangle,
\end{equation}
where $\mathcal{O}(x)$ is a local Hermitian operator and
\begin{equation}
	\mathcal{O}(x,t) = e^{i Ht - i P x}\mathcal{O}(0) e^{-i Ht + i Px}.
\end{equation}

In this work our main focus lies on correlation functions of the local particle density $\hat{q}_0(x) (\equiv \hat{\rho}(x))$, the associated particle current $\hat{j}_0(x)$ and their higher "spin" generalizations $\hat{q}_j(x,t)$ and $\hat{j}_j(x,t)$ present in the integrable theories. We introduce the following notation
\begin{align}
	C_{ij}(x,t) = \langle \vartheta| \hat{q}_i(x,t) \hat{q}_j(0,0) |\vartheta\rangle, \quad \Gamma_{ij}(x,t) = \langle \vartheta| \hat{j}_i(x,t) \hat{j}_j(0,0) |\vartheta\rangle,
\end{align}
and reserve $S(x,t) (= C_{00}(x,t))$ for the special case of the density-density correlation function. The Fourier transform of $S(x,t)$, the dynamic structure factor (DSF), is important for the Bragg spectroscopy experiments with ultra-cold atoms\cite{2001_Brunello_PRA_64,2011_Fabbri_PRA_83,PhysRevA.91.043617,PhysRevLett.115.085301}. The correlations of higher conserved charges and currents are important for the construction of the Generalized Hydrodynamics~\cite{10.21468/SciPostPhys.6.4.049,2018PhRvL.121p0603D}. 
The charge and current operators obey the continuity equation
\begin{equation}
	\partial_t \hat{q}_j(x,t) + \partial_x \hat{j}_j(x,t) = 0,
\end{equation}
which implies that the associated charge
\begin{equation}
	Q_j = \int {\rm d}x\, \hat{q}_j(x,t),
\end{equation}
is conserved. The expectation value of the charge $\hat{q}_j$ on the thermodynamic state $|\vartheta \rangle$ is
\begin{equation}
	\langle \vartheta |\hat{q}_j |\vartheta\rangle = \int {\rm d}\lambda\, \vartheta(\lambda) h_j(\lambda),
\end{equation}
where $h_j(\lambda)$ is the single particle eigenvalue with $h_0(\lambda) = 1$ for the density, $h_1(\lambda) = \lambda$ for the momentum and $h_2(\lambda) = \lambda^2$ for the energy of particles.

Local operators, like $\hat{q}_j(x)$, are not capable of modifying the thermodynamic state macroscopically, their action connects different microscopic realizations of the same state. In this work we are concerned with operators conserving the number of particles. Such operators connect states with the same number of particles. The relevant states are then parametrized as $m$ particles and holes modifying the filling function \begin{equation}
	\vartheta(\lambda; \bfp, \bfh) = \vartheta(\lambda) + \frac{1}{L} \sum_{j=1}^{m} \left( \delta(\lambda - p_j) - \delta(\lambda - h_j) \right),
\end{equation}
with respect to a chosen reference state $|\vartheta\rangle$. We denote the corresponding state $|\vartheta, \bfh, \bfp\rangle$. The energy and momentum of such excited states, with respect to $|\vartheta\rangle$~are
\begin{align}
	\omega(\bfp, \bfh) = \sum_{j=1}^m \left(\omega(p_j) - \omega(h_j)\right), \qquad k(\bfp, \bfh) = \sum_{j=1}^m \left(k(p_j) - k(h_j)\right), \label{kinematics}
\end{align}
where 
\begin{align}
	\omega(\lambda) = \lambda^2 + 2 \int {\rm d}\mu\, \vartheta(\mu) \mu F(\mu| \lambda), \qquad
	k(\lambda) = \lambda + \int {\rm d}\mu\, \vartheta(\mu) F(\mu| \lambda), 
\end{align}
with the back-flow function $F(\mu, \lambda)$ satisfying
\begin{align}
	F(\mu| \lambda) = \frac{\theta(\mu - \lambda)}{2\pi} + \int {\rm d}\mu' \vartheta(\mu') T(\mu, \mu') F(\mu'|\lambda).
\end{align} 
The phase shift $\theta(\lambda) = 2\, \atan(\lambda/c)$ and is related to the differential phase shift,
\begin{equation}
	T(\lambda, \lambda') = \frac{1}{2\pi} \frac{\partial \theta(\lambda - \lambda')}{\partial \lambda}.
\end{equation}
  
After characterizing the relevant excited states, we turn our attention to the correlation functions and write the resolution of the identity, in the subspace of the Hilbert space with a fixed number of particles, as
\begin{equation}
	\mathbf{1} = \sum_{m=0}^{\infty} \frac{1}{(m!)^2} \fint {\rm d}\bfp_m {\rm d}\bfh_m\, |\vartheta, \bfh, \bfp\rangle \langle \vartheta, \bfh, \bfp\rangle, 
\end{equation}
with the integration measure defined as
\begin{equation}
  {\rm d}\mathbf{p}_m {\rm d}\mathbf{h}_m = \prod_{j=1}^m {\rm d}p_j \,{\rm d}h_j \,\rho_h(p_j) \rho(h_j).
\end{equation}
and Hadamard regularization which excludes contributions when any particle $p_j$ coincides with any hole $h_k$. For example, the connected density-density correlation function in the spectral representation is 
\begin{equation}
	S(x,t) = \sum_{m=1}^{\infty} \frac{1}{(m!)^2}  \fint {\rm d}\bfp_m {\rm d} \bfh_m e^{i k(\bfp, \bfh) x - i \omega(\bfp, \bfh) t} |\langle \vartheta| \hat{\rho}(0) |\vartheta , \bfp, \bfh\rangle|^2. \label{corr_x-t}
\end{equation}
In this work, we focus on their Fourier transform
\begin{equation}
	S(k, \omega) = \int {\rm d}t  {\rm d}x \,e^{-ikx + i \omega t} S(x,t),
	\label{corr-k-omega}
\end{equation}
Contributions to the both expressions~\eqref{corr_x-t} and~\eqref{corr-k-omega} can be readily classified by the number of particle-hole pairs involved. For $S(k,\omega)$ we  write
\begin{equation}
	S(k,\omega) = \sum_{m=1}^\infty S^{m{\rm ph}}(k,\omega),
\end{equation}
where, as advertised in the introduction,
\begin{equation}
	S^{m{\rm ph}}(k,\omega) =  \frac{(2\pi)^2}{(m!)^2} \fint {\rm d}\bfp_m {\rm d} \bfh_m |\langle \vartheta| \hat{\rho}(0) |\vartheta , \bfp, \bfh\rangle|^2 \delta( k - k(\bfp,\bfh)) \delta(\omega - \omega(\bfp, \bfh)). \label{spectral}
\end{equation}
The expansion in the number of particle-hole excitations is also an expansion in the momentum where the $m{\rm ph}$ contribution is of order $k^{m-1}$~\cite{2018JSMTE..03.3102D}. The full correlation function at a given order in momentum $k$ contains contributions from excited states with different number of particle-hole excitations. The order $k^{-1}$ comes solely from the $\oneph$ excitations, but the order $k^0$ contains two contributions: the next-to-leading contribution from $\oneph$ excitations and the leading contribution from $\twoph$ excitations. The main aim of this work is to understand the latter contribution, that is the small momentum contribution to $S^{\twoph}(k, \omega)$ as this is the simplest instance in which the Hadamard regularization appears.

The thermodynamic form factors $\langle \vartheta| \mathcal{O}(0) |\vartheta , \bfp, \bfh\rangle$ can be computed either directly from a microscopic theory in the case where the microscopic form-factors are known, for example due to the Algebraic Bethe Ansatz~\cite{1990_Slavnov_TMP_82,KorepinBOOK,1997CMaPh.188..657K} or the vacuum Form-Factors bootstrap program~\cite{doi:10.1142/1115}. An alternative and direct route is provided by Thermodynamic Bootstrap Program~\cite{Bootstrap_JHEP,Cortes_Cubero_2020}.

The full thermodynamic form-factor of the density operator in the Lieb-Liniger model was computed in~\cite{Smooth_us}. Its small momentum limit was explored in~\cite{SciPostPhys.1.2.015,2018JSMTE..03.3102D}. Independently, small momentum limits of thermodynamic form-factors for higher conserved charges and currents and for 1 particle-hole excited state were investigated in~\cite{Doyon_drude,2018ScPP....5...54D}. The universal form of the 2 particle-hole thermodynamic form factors were conjectured in~\cite{10.21468/SciPostPhys.6.4.049,2018PhRvL.121p0603D}. The same structure arises in the Thermodynamic Bootstrap Program~\cite{Bootstrap_JHEP,Cortes_Cubero_2020} as the consequence of the annihilation pole axiom. From this point of view it is also possible to conjecture analogous expressions for higher particle-hole form-factors. In the following we summarize these findings, focusing on the leading contributions to the form-factors.

The one particle-hole form-factor of the conserved charge density is
\begin{equation}
	\langle\vartheta|\hat{q}_j|\vartheta, h,p \rangle = h_j^{\rm dr}(h) + (\dots). \label{oneph}
\end{equation}
The two particle-hole form-factor is
\begin{align}
	\langle\vartheta|\hat{q}_j(0)|\vartheta, \bfh_2, \bfp_2 \rangle =&\, 2\pi  k(\bfp, \bfh) \times \nonumber \\ 
	&\left(\frac{T^{\rm dr}(h_2, h_1) h_j^{\rm dr}(h_2) }{k'(h_1) k'(h_2) (p_1 - h_1)}+ \frac{T^{\rm dr}(h_1, h_2) h_j^{\rm dr}(h_1) }{k'(h_2) k'(h_1) (p_2 - h_2)} \right. \nonumber\\
	&\left. + \frac{T^{\rm dr}(h_2, h_1) h_j^{\rm dr}(h_2) }{k'(h_1) k'(h_2) (p_2 - h_1)} + \frac{T^{\rm dr}(h_1, h_2) h_j^{\rm dr}(h_1) }{k'(h_2) k'(h_1) (p_1 - h_2)} + (\dots) \right). \label{twoph}
\end{align}
Combining conjectures from~\cite{10.21468/SciPostPhys.6.4.049} and~\cite{Bootstrap_JHEP} we can guess the structure of higher form-factors. For example, the next one is
\begin{align}
	\langle\vartheta|\hat{q}_j(0)|\vartheta, \bfh_3, \bfp_3\rangle =&\, \frac{(2\pi k(\bfp, \bfh))^2}{k'(h_1) k'(h_2) k'(h_3)} \times \nonumber \\
	&\left( \frac{T^{\rm dr}(h_1, h_3) T^{\rm dr}(h_2, h_3) h_j^{\rm dr}(h_3)}{(p_1 - h_1)(p_2 - h_2)} + (\text{cycl. perm.}) + (\dots)\right), \label{threeph}
\end{align}
where $(\text{cycl. perm.})$ refers to independent cyclic permutations of indices of particles and holes. The reasoning leading to this last formula is presented in Appendix~\ref{app:3ph}. In all these expressions $(\dots)$ stands for subleading contributions in the particle-hole differences $p_i - h_j$. 

The index ${\rm dr}$ appearing in the formulas above describes the dressing procedure: $f(\lambda) \overset{\rm dressing}{\longrightarrow} f^{\rm dr}(\lambda)$, defined as
\begin{equation}
	f^{\rm dr}(\lambda) = f(\lambda) + \int {\rm d}\lambda'\, \vartheta(\lambda') T(\lambda, \lambda') f^{\rm dr}(\lambda').
\end{equation}
The dressing is relative to the filling function $\vartheta(\lambda)$. The (left-)dressing of the differential scattering kernel $T(\lambda, \lambda')$ is 
\begin{equation}
	T^{\rm dr}(\lambda, \lambda') = T(\lambda, \lambda') + \int {\rm d}\lambda'' \vartheta(\lambda'') T(\lambda, \lambda'') T^{\rm dr}(\lambda'', \lambda'),
\end{equation}
with the resulting function inheriting the symmetry in exchanging the arguments. 

The filling function $\vartheta(\lambda)$ characterizes the state of the system. At (generalized) thermal equilibrium it is given by
\begin{equation}
	\vartheta(\lambda) = \frac{1}{1 + e^{\epsilon(\lambda)}}, \label{thermal_vartheta}
\end{equation}
with $\epsilon(\lambda)$ solving the generalized TBA equation~\cite{1969_Yang_JMP_10,2012_Mossel_JPA_45,2012PhRvL.109q5301C}
\begin{equation}
	\epsilon(\lambda) = \sum_{j} \mu_j h_j(\lambda) - \mu - \frac{1}{2\pi} \int {\rm d}\lambda' \,K(\lambda - \lambda') \log\left( 1 + e^{-\epsilon(\lambda')}\right).
\end{equation}
The case of a thermal gas at temperature $T$ is given by $\mu_2 = 1/T$ (we set $k_B = 1$) with other chemical potentials $\mu_j$ equal to zero. 

The thermodynamic form-factors are known for smooth filling function $\vartheta(\theta)$, the thermal one~\eqref{thermal_vartheta} being an important example. However, once the correlation function is computed it is possible to take the limit to discontinuous filling function, for example the one describing the ground state~\cite{1963_Lieb_PR_130_1},
\begin{equation}	
	\vartheta(\lambda) = 1 - \Theta(|\lambda| - \lambda_F), \label{gs_vartheta}
\end{equation}
with $\lambda_F$ the Fermi rapidity set by the density of particles $N/L$.

\section{Small excitations and the linearization of the dressed momentum and energy} \label{sec:small}

In eqs.~\eqref{oneph}, \eqref{twoph} and \eqref{threeph}, the most singular part of the form-factors in $p_i -h_j$ is shown. It forms the leading part of the form-factor whenever particles and holes can be paired such that for each pair $(p_i,h_i)$ the difference $p_i - h_i$ is small. We call such excitations \emph{small particle-hole excitations}. In this section we formalize this concept by i) introducing a criteria saying when $p_i -h_i$ is small and ii) exploring a relation between states formed with small excitations and their momentum and energy.

Let us start with a single particle-hole excited state $|\vartheta, h, p\rangle$ specified by $(p,h)$. Its kinematics, according to~\eqref{kinematics}, is 
\begin{align}
	k = k(p) - k(h), \qquad
	\omega = \omega(p) - \omega(h). \label{kinematics_1ph}
\end{align}
When $p-h$ is small, given the analyticity of $k(\lambda)$ and $\omega(\lambda)$, we can expand both formulas to~find
\begin{align}
	k &= k'(h) (p-h) + \frac{1}{2} k''(h) (p-h)^2 + \mathcal{O}\left( (p-h)^3 \right), \\
	\omega &= \omega'(h) (p-h) + \frac{1}{2} \omega''(h) (p-h)^2 + \mathcal{O}\left( (p-h)^3 \right).
\end{align}
The small excitations assumption breaks certainly when the first and second order terms in these expansions become similar. These leads to the constraints
\begin{equation}
	\lvert p-h \rvert \lesssim 2 \left\lvert \frac{k'(h)}{k''(h)} \right\vert, \qquad \lvert p-h \rvert \lesssim 2 \left\lvert \frac{\omega'(h)}{\omega''(h)} \right\rvert,
\end{equation}
which, when fulfilled, allow for the linearization of the dressed momentum and energy. From numerical solutions we observe that the second condition is tighter and implies the first\footnote{This is partially because functions $k'(\lambda)$ and $\omega''(\lambda)$ are bounded from below, whereas $k''(\lambda)$ and $\omega'(\lambda)$ are odd functions of the rapidity. This makes, at least at small $h$, the momentum bound to diverge and the energy bound to fall to zero and therefore making it tighter.}. This leads to the following condition
\begin{equation}
\lvert p-h \rvert < 2 \left\lvert \frac{\omega'(h)}{\omega''(h)} \right\rvert. \label{energy_bound}
\end{equation}
Such excitation has then a linear spectrum $\omega = v^{\rm eff}(h) k$, with the velocity dependent on its position in the rapidity space. Alternatively, we can think of $h$ as labelling different particles types, each with a different velocity of propagation. The effective velocity is defined as
\begin{equation}
	v^{\rm eff}(\lambda) = \frac{\omega'(\lambda)}{k'(\lambda)}.
\end{equation}

\begin{figure}
	\center
	\includegraphics[scale=0.75]{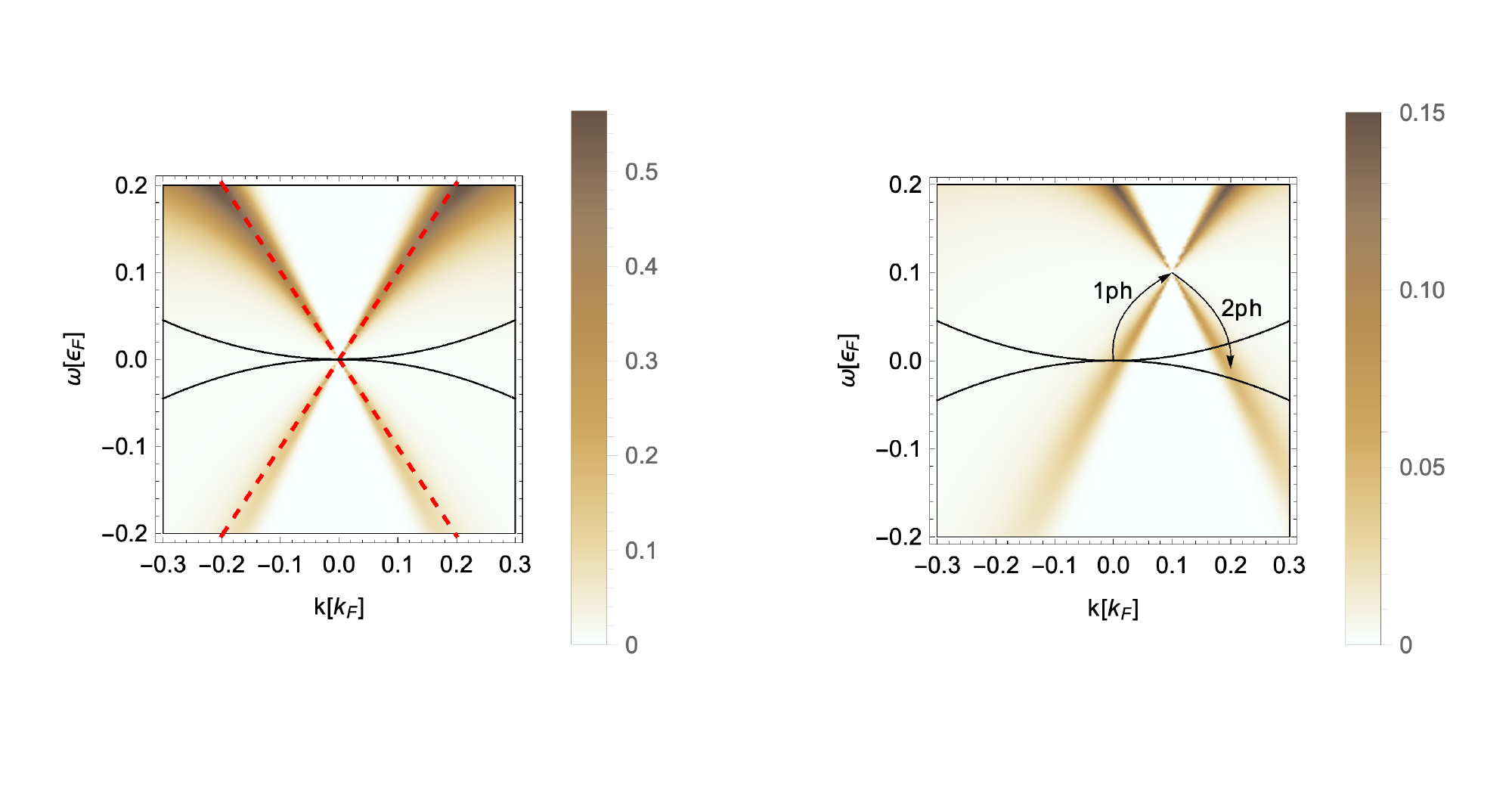}
	\caption{(\emph{left panel:}) The phase space for small $\oneph$ excitations. The color codes for the phase space factor $\vartheta(h)(1 - \vartheta(p))$ with $(p,h)$ solving~\eqref{kinematics_1ph} at the linear order in $p-h$, the black line is the bound~\eqref{energy_bound} and the dotted red lines show the dispersion relation for the excitations with velocity $\bar{v}^{\rm eff}$. (\emph{right panel:}) The same but now for $\twoph$ excitations with the first excitation fixed as shown by the $\oneph$ arrow. The color codes for the phase factor $\vartheta(h_1)(1 - \vartheta(p_1))\vartheta(h_2)(1 - \vartheta(p_2))$. Both plots are for the thermal state with $T=1$, $c=4$ and in units of Fermi momentum $k_F = \pi N/L$ and Fermi energy $\epsilon_F = k_F^2$.} 
	\label{fig:phExcPlot}
\end{figure}

We can now check for what values of $k$ and $\omega$ the small excitations exist. That is, we fix $k$ and $\omega$, find the corresponding $p$ and $h$ at the linear order, and then check if their difference satisfies the bound~\eqref{energy_bound}. Additionally, to get an impression which excitations are important we look at the value of $\vartheta(h)(1 - \vartheta(p))$ which controls how much space there is to create the excitation. The results are shown in fig.~\ref{fig:phExcPlot}. We observe numerically that the possible excitations are localized along two rays determined by the average velocity 
\begin{equation}
	\bar{v}^{\rm eff} = \frac{\int {\rm d}h \,\vartheta(h) (1 - \vartheta(h)) v^{\rm eff}(h)}{\int {\rm d}h \,\vartheta(h) (1 - \vartheta(h))}.
\end{equation}

The excluded region is a region with energy $\omega$ relatively small compared to $k \bar{v}^{\rm eff}$ and corresponds to excitations localized in the center of the distribution $\vartheta(\lambda)$, that is with $h$ and $p$ close to $0$. The energy $\omega(\lambda)$ has a global minimum in this region making breaking the bound easy. On the other hand for the thermal states of not too high temperature, the centre of distribution $\vartheta(\lambda)$ is largely filled which effectively diminishes their contributions to the correlation functions. This shows, that the relevant part of the $\oneph$ excited states falls into the class of small particle-hole excited states for which the dressed momentum and energy can be linearized.

The situation gets only slightly more complicated for the two particle-hole excitations. We want both excitations to constitute small excitations, that is 
\begin{equation}
	\lvert p_i-h_i \rvert \lesssim 2 \left\lvert \frac{\omega'(h_i)}{\omega''(h_i)} \right\rvert, \qquad i=1,2.
\end{equation}
The phase space is then constructed in the standard way: each point on the single ph phase diagram becomes now a point from which rays of the second ph excitation originate, see fig.~\ref{fig:phExcPlot}. In the consequence, the $\twoph$ excitations are not focused only along the two rays, but rather cover the whole plane of small $k-\omega$ including the $\oneph$ excluded region.

As we are interested in correlation functions at small momentum and energy it is important to understand whether there are  small momentum-energy excited states that escape this structure. For the energy and momentum of the excited states to be small, in the leading order, we have relations
\begin{equation}
	k(p_1) + k(p_2) \approx k(h_1) + k(h_2), \qquad
	\omega(p_1) + \omega(p_2) \approx \omega(h_1) + \omega(h_2).
\end{equation}
One can interpret this condition from the point of view of an elastic scattering process in $(1+1)d$ with initial rapidities $p_1$ and $p_2$ and final rapidities $h_1$ and $h_2$. The sets of initial and final rapidities must then coincide up to a permutation
\begin{equation}
	\begin{cases}
		p_1 \approx h_1, \\
		p_1 \approx h_2,
	\end{cases}
	\qquad
	\begin{cases}
		p_1 \approx h_2 \\
		p_2 \approx h_1,
	\end{cases}
\end{equation}
with $h_1$ and $h_2$ free parameters. Whether the energy bound~\eqref{energy_bound} is fulfilled depends then on their values. In any case, there are choices of $h_1$ and $h_2$ such that both pairs constitute small excitations. Such choices are again dominating by the phase space argument.
Therefore, the relevant $\twoph$ excited states are formed by the small particle-hole excitations for which again the dressed momentum and energy can be linearized.

Considering $3{\rm ph}$ and higher excited states the momentum and energy constraints, together with the phase space argument, are not enough to restrict the structure of the excited state to be of small excitations type. 

In summary, we have argued that small excitations reign the contributions of $\oneph$ and $\twoph$ excitations to the DSF in the small momentum-energy limit. Each small ph excitation has a linear dispersion relation with possibly different velocity, however for thermal states focused around~$\bar{v}^{\rm eff}$. 

\section{2ph contribution to the DSF} \label{sec:2ph}

In this section we evaluate the contribution to the DSF from small $\twoph$ excitations. For the completeness, in Appendix~\ref{app:1ph}, we have included the computations in the $\oneph$ case. We start by specializing eq.~\eqref{twoph} to the density form-factor, using $h_0^{\rm dr}(\lambda) = 2\pi \rho_t(\lambda)$, 
\begin{align}
	\langle\vartheta|\hat{q}_j(0)|\vartheta, \bfh_2, \bfp_2 \rangle =&\, (2\pi)^2  k(\bfp, \bfh) \times \nonumber \\ 
	&\left(\frac{T^{\rm dr}(h_2, h_1) \rho_t(h_2) }{k'(h_1) k'(h_2) (p_1 - h_1)}+ \frac{T^{\rm dr}(h_1, h_2) \rho_t(h_1) }{k'(h_2) k'(h_1) (p_2 - h_2)} \right. \nonumber\\
	&\left. + \frac{T^{\rm dr}(h_2, h_1) \rho_t(h_2) }{k'(h_1) k'(h_2) (p_2 - h_1)} + \frac{T^{\rm dr}(h_1, h_2) \rho_t(h_1) }{k'(h_2) k'(h_1) (p_1 - h_2)} + (\dots) \right). \label{2ph_sec4}
\end{align}
The kinematics~\eqref{kinematics}, for two small particle-hole excitations, is
\begin{align}
	k(\mathbf{p}, \mathbf{h}) &= k'(h_1) (p_1 - h_1) + k'(h_2) (p_2 - h_2), \nonumber\\ 
	\omega(\mathbf{p}, \mathbf{h}) &= v^{\rm eff}(h_1) k'(h_1) (p_1 - h_1) + v^{\rm eff}(h_2) k'(h_2) (p_2 - h_2).
\end{align}
Here, we have implicitly assumed a pairing ($p_1$ with $h_1$ and $p_2$ with $h_2$) of $\bfp$ with $\bfh$. Given the symmetry of the form-factor, to account for the other choice it suffices to multiply the contribution to the DSF by $2$. 
It is convenient to change the variables from $(\bfp, \bfh)$ to $(\bfa, \bfh)$ where $\bfa = \bfp - \bfh$. 
The contribution~\eqref{spectral} to the DSF is then
\begin{equation}
	S^{\twoph}(k,\omega) = \frac{(2\pi)^2}{2} \int {\rm d}h_1 {\rm d}h_2 \fint_0 {\rm d}\alpha_1 {\rm d}\alpha_2 F(\bfh, \bfa) \delta( k - k( \bfa + \bfh, \bfh)) \delta( \omega - \omega( \bfa + \bfh, \bfh)), \label{foo}
\end{equation}
where
\begin{equation}
	F(\bfh, \bfa) = \rho_p(h_1) \rho_p(h_2) \rho_h(h_1 + \alpha_1) \rho_h(h_2 + \alpha_2) |\langle\vartheta|\hat{\rho}(0)|\vartheta, \mathbf{h},\bfa + \bfh\rangle|^2.
\end{equation}
We will now perform integrations over $\bfa$ with the help of the $\delta$-functions. Certain care is required due to the Hadamard regularization. At first, let us proceed as if these were regular integrals. The presence of the $\delta$-function puts the form-factor on the $(k,\omega)$-shell. That is, given the momentum and energy constraints: $k = k(\bfa+\bfh, \bfh)$ and $\omega = \omega(\bfa + \bfh, \bfh)$, we can solve for the differences $\alpha_i$ to find
\begin{align} \label{alphabar}
\begin{split}
	\bar{\alpha}_1 =  \frac{\omega - v^{\rm eff}(h_2) k}{k'(h_1)(v^{\rm eff}(h_1) - v^{\rm eff}(h_2))}, \\
	\bar{\alpha}_2 = \frac{\omega - v^{\rm eff}(h_1) k}{k'(h_2)(v^{\rm eff}(h_2) - v^{\rm eff}(h_1))}.
\end{split}
\end{align}
The product of the $\delta$-functions can be now disentangled giving 
\begin{equation}
	\delta( k - k( \bfa + \bfh, \bfh)) \delta( \omega - \epsilon( \bfa + \bfh, \bfh)) = 
	\frac{\delta(\alpha_1 - \bar{\alpha}_1) \delta(\alpha_2 - \bar{\alpha}_2)}{k'(h_1) k'(h_2) |v^{\rm eff}(h_1) - v^{\rm eff}(h_2)|}.
\end{equation} 
This yields
\begin{equation}
	F(\bfh, \bfa) \delta( k - k( \bfa + \bfh, \bfh)) \delta( \omega - \epsilon( \bfa + \bfh, \bfh)) = F_{\bfh}(k,\omega) \delta(\alpha_1 - \bar{\alpha}_1) \delta(\alpha_2 - \bar{\alpha}_2),
\end{equation}
where $F_{\bfh}(k,\omega)$ is the $(k,\omega)$-shell form-factor together with the density factors and extra contributions from the integrals over $\delta$-functions
\begin{align}
	F_{\bfh}(k,\omega) &= \frac{F(\bfh, \bar{\bfa})}{k'(h_1) k'(h_2) |v^{\rm eff}(h_1) - v^{\rm eff}(h_2)|} \nonumber \\
	& = \vartheta(h_1) \vartheta(h_2) \rho_h(h_1 + \alpha_1) \rho_h(h_2 + \alpha_2) \frac{|\langle\vartheta|\hat{\rho}(0)|\vartheta, \mathbf{h},\bar{\bfa} + \bfh\rangle|^2}{(2\pi)^2 |v^{\rm eff}(h_1) - v^{\rm eff}(h_2)|}. \label{ko_ff}
\end{align}
Ignoring the Hadamard regularization in~\eqref{foo} we then find
\begin{equation}
	S^{\twoph}(k,\omega) = \frac{(2\pi)^2}{2} \int {\rm d}h_1 {\rm d}h_2 F_{\bfh}(k,\omega).
\end{equation}
Let us look into the structure of $F_{\bfh}(k,\omega)$ and start with evaluating the form-factor $\langle\vartheta|\hat{\rho}(0)|\vartheta, \bfh, \bfp\rangle$ with positions of the particles $\bfp = \bfh + \bar{\bfa}$ fixed by the momentum and energy $(k,\omega)$. Recall from~\eqref{2ph_sec4}, that the form factor consists of $4$ terms potentially diverging like $p_i - h_j$ and some regular parts. Additionally, the form-factor is multiplied by the total momentum. In the small momentum limit only the diverging parts of the form-factor contribute as they are also of order of $k$. For the choice of the excited state, the terms containing $p_1-h_1$ and $p_2 - h_2$ are small and therefore constitute the leading contribution. The form-factor can be simplified to
\begin{align}
		\langle\vartheta|\hat{\rho}(0)|\vartheta, \bfh, \bfp\rangle = (2\pi)^2 k(\bfh, \bfp) 
	\left(\frac{T^{\rm dr}(h_2, h_1) \rho_t^{\rm dr}(h_2) }{k'(h_1) k'(h_2) (p_1 - h_1)}+ \frac{T^{\rm dr}(h_1, h_2) \rho_t^{\rm dr}(h_1) }{k'(h_2) k'(h_1) (p_2 - h_2)} \right).
\end{align}
Using that $T^{\rm dr}(h_1, h_2) = T^{\rm dr}(h_2, h_1)$ and $k'(h) = 2\pi \rho_t(h)$ together with the definition of $\bfa$, the expression simplifies further to 
\begin{equation}
	\langle\vartheta|\hat{\rho}(0)|\vartheta, \bfh, \bfp\rangle = 2 \pi k(\bfh, \bfp) T^{\rm dr}(h_1, h_2)
	\frac{k'(h_1) \alpha_1 + k'(h_2) \alpha_2}{k'(h_1)k'(h_2) \alpha_1 \alpha_2} .
\end{equation}
Putting the form-factor on $(k,\omega)$-shell means setting $\alpha_i = \bar{\alpha}_i$. In that case  we find
\begin{align}	
	\langle\vartheta|\hat{\rho}(0)|\vartheta, \bfh, \bar{\bfp}\rangle
	= - 2\pi T^{\rm dr}(h_1, h_2) \frac{k^2(v^{\rm eff}(h_1) - v^{\rm eff}(h_2))^2}{(\omega - v^{\rm eff}(h_1)k)(\omega - v^{\rm eff}(h_2)k)}.
\end{align}	
We see now the problem with ignoring the Hadamard regularization. The Hadamard regularization takes care of the unphysical divergence when particle and hole are placed on top of each other. This corresponds to the pole in the form-factor when $\alpha_i \approx 0$. The $(k,\omega)$-shell form-factor still has the same pole, it appears now as a pole when $\omega - v^{\rm eff}(h_1) k \approx 0$. This happens, when one of the particle-hole excitations carries the whole energy and momentum of the excited state, meaning, that for the other excitation, $\alpha_i \approx 0$. By properly executing the Hadamard regularized integrals we should then find the integrations over $h_i$ also properly regularized. The correct answer turns out to be simply
\begin{equation}
	S^{\twoph}(k,\omega) = \frac{(2\pi)^2}{2}\fint_{h^*} {\rm d}h_1{\rm d} h_2\, F_{\bfh}(k,\omega),
\end{equation}
with the double Hadamard integral. Parameter $h^*$ is defined as the unique solution to $\omega = v^{\rm eff}(h^*) k$. The uniqueness comes from the fact that $v^{\rm eff}(h^*)$ is a strictly monotonic function. In the remaining part of this section we present the derivation of this result.

\subsection{Derivation of the main result}

The Hadamard regularization assigns a finite value to an integral of a function with a double pole. We need to evaluate two Hadamard integrals of a function with the pole structure $1/|\alpha_1 \alpha_2|^2 $. 
Recall the Hadamard regularization for a generic function $f(x)$ having a double pole at $x=0$,
\begin{equation}
	\fint_0 {\rm d}x\, f(x) = \lim_{\epsilon \rightarrow 0^+} \int {\rm d}x\,  \left( f(x) \Theta(|x| - \epsilon) - \frac{2}{\epsilon} \delta( x) f^{(1)}(x) \right), \label{Hadamard_prescription}
\end{equation}
where $f^{(1)}(x) \equiv x^2 f(x)$ and which has now a finite value for $x=0.$\footnote{In practice to evaluate the Hadamard regularized integral we need only $f^{(1)}(x)$ at $x=0$. We keep the $x$-dependence as it simplifies the notation in the $2d$ case.}
We introduce the following notation for the form-factors
\begin{equation}
	F^{(i)}(\bfh, \bfa) = \alpha_i^2 F(\bfh, \bfa), \qquad F^{(1,2)}(\bfh, \bfa) = \alpha_1^2 \alpha_2^2 F(\bfh, \bfa),
\end{equation}
and for their $(k,\omega)$-shell counterparts
\begin{align}
	F_{\bfh}^{(i)}(k, \omega) = \bar{\alpha}_i^2 F_{\bfh}(k, \omega), \qquad 
	F_{\bfh}^{(1,2)}(k, \omega) = \bar{\alpha}_1^2 \bar{\alpha}_2^2 F_{\bfh}(k, \omega),
\end{align}
and consider
\begin{equation}
	F_{\bfh} (k, \omega) = \fint_0 {\rm d}\alpha_1 {\rm d}\alpha_2 F(\bfh, \bfa) \delta( k - k(\bfa + \bfh, \bfh)) \delta(\omega - \epsilon(\bfa + \bfh, \bfh)).
\end{equation}
Evaluating the two integrals, according to the prescription~\eqref{Hadamard_prescription}, leads to $3$ different contributions, classified by the number of $\epsilon_i$'s multiplying the contribution: zero, one or two. The first one (without $\epsilon_i$ prefactor) is
\begin{align}
	&\int {\rm d}\alpha_1 {\rm d}\alpha_2 F(\bfh, \bfa) \delta( k - k(\bfa + \bfh, \bfh)) \delta(\omega - \epsilon(\bfa + \bfh, \bfh)) \Theta(|\alpha_1| - \epsilon_1) \Theta(|\alpha_2| - \epsilon_2) \nonumber \\
	& = F_{\bfh}(k,\omega) \Theta(|\bar{\alpha}_1| - \epsilon_1) \Theta(|\bar{\alpha}_2| - \epsilon_2). \label{Hadamard_one}
\end{align}
There are two contributions of the second type. One of the form
\begin{align}
	& - \frac{2}{\epsilon_1} \int {\rm d}\alpha_1 {\rm d}\alpha_2 F^{(1)}(\bfh, \bfa) \delta( k - k(\bfa + \bfh, \bfh)) \delta(\omega - \epsilon(\bfa + \bfh, \bfh))  \delta(\alpha_1)\Theta(|\alpha_2| - \epsilon_2) \nonumber \\
	&=  -\frac{2}{\epsilon_1} F^{(1)}_{\bfh}(k, \omega) \delta(\bar{\alpha}_1)  \Theta(|\bar{\alpha}_2| - \epsilon_2), \label{Hadamard_two}
\end{align}
and the second with indices $1$ and $2$ exchanged. Finally, the last contribution, with both $\epsilon_i$'s in the prefactor, is
\begin{align}
	&  \frac{4}{\epsilon_1 \epsilon_2} \int {\rm d}\alpha_1 {\rm d}\alpha_2 F^{(1,2)}(\bfh, \bfa) \delta( k - k(\bfa + \bfh, \bfh)) \delta(\omega - \epsilon(\bfa + \bfh, \bfh))  \delta(\alpha_1) \delta(\alpha_2) \nonumber \\
	&=  \frac{4}{\epsilon_1 \epsilon_2} F^{(1,2)}_{\bfh}(k, \omega) \delta(\bar{\alpha}_1) \delta(\bar{\alpha}_2), \label{Hadamard_three}
\end{align}
Summing all the contributions we find
\begin{align}
	F_{\bfh}(k, \omega) = \lim_{\epsilon_i \rightarrow 0} &\left(F_{\bfh}(k,\omega) \Theta(|\bar{\alpha}_1| - \epsilon_1) \Theta(|\bar{\alpha}_2|- \epsilon_2)  - \frac{2}{\epsilon_1} F^{(1)}_{\bfh}(k, \omega) \delta(\bar{\alpha}_1)  \Theta(|\bar{\alpha}_2| -\epsilon_2)\right. \nonumber \\
	&\left. -  \frac{2}{\epsilon_2} F^{(2)}_{\bfh}(k, \omega)  \Theta(|\bar{\alpha}_1|-\epsilon_2) \delta(\bar{\alpha}_2) + \frac{4}{\epsilon_1 \epsilon_2} F^{(1,2)}_{\bfh}(k, \omega) \delta(\bar{\alpha}_1) \delta(\bar{\alpha}_2)\right). \label{Hadamard_sum}
\end{align}
For a given value of $k$ and $\omega$ this expression should be integrated over $h_1$ and $h_2$ and then the limits $\epsilon_i \rightarrow 0$ should be taken. This is still quite complicated because $\bar{\alpha}_i$ depends on both $h_1$ and $h_2$. We would like to rewrite this expression in such a way, that the conditions on $\bfh$ are explicit. This is easy to achieve for the first term of~\eqref{Hadamard_sum}, where we can use two properties of the Heaviside $\Theta$-function,
\begin{equation}
	\Theta(x/a) = \Theta(x), \qquad \Theta(f(x)) = \Theta(x).
\end{equation}
The first one holds for $a>0$, while the second one for an odd function $f(x)$ with positive values for $x>0$. We have then a chain of relations
\begin{equation}
	\Theta(|\bar{\alpha}_1| - \epsilon_1) = \Theta(|v^{\rm eff}(h^*) - v^{\rm eff}(h_2)|  - \epsilon_1') = \Theta (|h^* - h_2| - \epsilon_1''),
\end{equation} 
with new $\epsilon_1' = |k'(h_1)(v^{\rm eff}(h_1) - v^{\rm eff}(h_2))| \epsilon_1 $ and $\epsilon_1'' = \epsilon_1' / | k\kappa(h^*)|$. Here $\kappa(\lambda) = \partial_{\lambda} v^{\rm eff}(\lambda)$ and we used that $v^{\rm eff}(\lambda)$ is a monotonically increasing function. We can now perform similar transformations with other expressions appearing in~\eqref{Hadamard_sum}. For example,
\begin{equation}
	\delta(\bar{\alpha}_1) = \frac{\epsilon_1''}{\epsilon_1} \delta(h^* - h_2).
\end{equation}
Transformation of $F_{\bfh}^{(i)}(k,\omega)$ is more subtle. Recall that its definition is the following: having function $f(x,y)$ with a double pole at $x$ we define $f^{(1)}(x) = x^2 f(x, y)$. $F_{\bfh}^{(1)}(k,\omega)$ is a function having a pole for $\alpha_1 = 0$. What we want is however to restate this in terms of the $h_2$ variable, with the pole now for $h_2 - h^* = 0$. In the vicinity of the pole we can then write 
\begin{equation}
	F_{\bfh}^{(1)}(k,\omega) = \left( \frac{\epsilon_1}{\epsilon_1''}\right)^2 \bar{F}_{\bfh}^{(2)}(k, \omega).
\end{equation}
with $\bar{F}_{\bfh}^{(2)}(k, \omega)$ defined as
\begin{equation}
	\bar{F}_{\bfh}^{(2)}(k, \omega) = (h_2 - h^*)^2 F_{\bfh}(k,\omega).
\end{equation}
Collecting together the rescalings of the Dirac $\delta$-function and of $F_{\bfh}^{(1)}(k,\omega)$ we find that the second term of~\eqref{Hadamard_sum} transforms as 
\begin{equation}
	- \frac{2}{\epsilon_1} F^{(1)}_{\bfh}(k, \omega) \delta(\bar{\alpha}_1)  \Theta(|\bar{\alpha}_2| -\epsilon_2) = - \frac{2}{\epsilon_1''} \bar{F}_{\bfh}^{(2)} \delta(h_2 - h^*) \Theta( |h_1 - h^*| - \epsilon_2'').
\end{equation}
Transforming in the similar way the remaining terms, dropping the bar from $\bar{F}_{\bfh}^{(i)}(k, \omega)$ and exchanging indices of $\epsilon_j$'s so they match indices of $h_j$'s, we find
\begin{align}
	F_{\bfh}(k, \omega) = \frac{1}{2}\lim_{\epsilon_i \rightarrow 0} &\left( F_{\bfh}(k,\omega) \Theta(|\Delta h_2^*| - \epsilon_2) \Theta(|\Delta h_1^*| - \epsilon_1) -\frac{2}{\epsilon_1}  F_{\bfh}^{(1)}(k, \omega) \delta(\Delta h_1^*)  \Theta(|\Delta h_2^*| - \epsilon_2)\right. \nonumber \\
	&\left. - \frac{2}{\epsilon_2} F_{\bfh}^{(2)}(k, \omega)\Theta(|\Delta h_1^*| - \epsilon_1) \delta(\Delta h_2^*) +  \frac{4}{\epsilon_1 \epsilon_2}  F_{\bfh}^{(1,2)}(k,\omega) \delta(\Delta h_1^*) \delta(\Delta h_2^*)\right)\!,
\end{align}
where $\Delta h_j^* = h_j - h^*$.
The integration over $F_{\bfh}(k, \omega)$ takes now a form of a double Hadamard integral of a function of two variables with a double pole in each variable at $h_i = h^*$. The $\twoph$ contribution to DSF then equals
\begin{equation}
	S^{\twoph}(k,\omega) = \frac{(2\pi)^2}{2}\fint_{h^*} {\rm d}h_1{\rm d} h_2\, F_{\bfh}(k,\omega). \label{twoph_DSF}
\end{equation}
with the following ingredients entering this formula
\begin{align}
	F_{\bfh}(k, \omega) &= G_{\bfh}(k, \omega) \times \frac{k^4 |v^{\rm eff}(h_1) - v^{\rm eff}(h_2)|^3}{(\omega - v^{\rm eff}(h_1)k)^2(\omega - v^{\rm eff}(h_2)k)^2}, \label{F_final} \\
	G_{\bfh}(k, \omega) &= \vartheta(h_1) \vartheta(h_2) \rho_h(h_1 + \bar{\alpha}_1) \rho_h(h_2 + \bar{\alpha}_2) \times \left( \frac{T^{\rm dr}(h_1, h_2)}{2\pi}\right)^2. \label{G_final}
\end{align}
The limiting expressions are
\begin{align}
	 F_{h^*, h_2}^{(1)} (k, \omega) &= \lim_{h_1 \rightarrow h^*} (h_1 - h^*)^2 F_{\bfh}(k, \omega) = G_{h^*, h_2}(k, \omega) \times \frac{|v^{\rm eff}(h_2) - v^{\rm eff}(h^*)|}{\kappa(h^*)^2}, \nonumber \\
	 F_{h_1, h^*}^{(1)} (k, \omega) &= \lim_{h_2 \rightarrow h^*} (h_2 - h^*)^2 F_{\bfh}(k, \omega) = G_{h_1, h^*}(k, \omega) \times \frac{|v^{\rm eff}(h_1) - v^{\rm eff}(h^*)|}{\kappa(h^*)^2},  \\
	 F_{h^*, h^*}^{(1,2)} (k, \omega) &= \lim_{h_1 \rightarrow h^*}  \lim_{h_2 \rightarrow h^*}  (h_1 - h^*)^2 (h_2 - h^*)^2 F_{\bfh}(k, \omega) = 0. \nonumber 
\end{align}
For smooth distributions of the rapidities and in the leading order in $k$, $G_{\bfh}(k, \omega)$ is further simplified by setting $\bar{\alpha}_i = 0$ in the arguments of $\rho_h(h)$. In that case $G_{\bfh}(k, \omega) = G_{\bfh}$ is a function of positions of the two holes only.

Eq.~\eqref{twoph_DSF} is the main result of this work. We will now turn to two applications of this formula. First, it should be clear that the derivation does not depend on the details of the form-factor considered, rather on its general analytic structure and therefore holds also for correlation functions of higher conserved charges and also of conserved currents. The computations are presented in Section~\ref{sec:higher}. Second, similarly as in the $\oneph$ case (see Appendix~\ref{app:1ph}), we can consider the ground state correlation function. This is presented in Section~\ref{sec:gs}. Before that, we discuss certain properties of the correlator.

\subsection{Properties of the correlator}

We close this section with a discussion of properties of $S^{\twoph}(k,\omega)$. To this end, we consider an alternative representation of $F_{\bfh}(k, \omega)$ which follows from parametrizing $\omega$ and $k$ with $h^*$, that is writing $\omega = v^{\rm eff}(h^*) k$. This leads to 
\begin{align}
	F_{\bfh}(h^*) &= G_{\bfh} \times \frac{|v^{\rm eff}(h_1) - v^{\rm eff}(h_2)|^3}{(v^{\rm eff}(h^*) - v^{\rm eff}(h_1))^2(v^{\rm eff}(h^*) - v^{\rm eff}(h_2))^2}.
\end{align}
This representation is convenient for considering various features of the correlator. For example, it is easy to see that $S^{\twoph}(k, \omega) = S^{\twoph}(k', \omega')$ if $\omega/k = \omega'/k'$ and as a consequence $S^{\twoph}(k,\omega) = S^{\twoph}(-k, -\omega)$. This is the low energy limit of the detailed balance relation, which in full generality reads
\begin{equation}
	S(k,\omega) = e^{-\omega/T} S(-k, -\omega).
\end{equation}
In the leading order in energy, the exponential term can be neglected leading to the aforementioned equality.

It is equally straightforward to show that $S^{\twoph}(k,\omega) = S^{\twoph}(-k,\omega)$. To this end we use that changing the sign of $k$ leads to a change in the sign of $h^*$ since $v^{\rm eff}(h)$ is an odd function. The change of sign of $h^*$ in the form-factor can be compensated by changing signs of the integration variables $h_1$ and $h_2$. This leads back to the original expression for the correlator upon assumption that the thermodynamic functions $\vartheta(h)$, $\rho_t(h)$ are even functions and $T^{\rm dr}(h_1, h_2) = T^{\rm dr}( -h_1, - h_2)$. This is the case at the thermal equilibrium. Therefore $S^{\twoph}(k,\omega) = S^{\twoph}(-k, \omega)$ and similarly $S^{\twoph}(k,\omega) = S^{\twoph}(k, -\omega)$. The first of these equalities reflects the spatial parity invariance of the thermal state. The second relation can be interpreted from the point of view of the f-sum rule, which reads
\begin{equation}
	\int \frac{{\rm d}\omega }{2\pi} \omega S(k,\omega) = \frac{N}{L} k^2.
\end{equation}
Given the aforementioned symmetry, the leading $\twoph$ contribution to this expression is zero.

Also of interest are the limiting expressions of zero momentum and energy. The result depends on the order of the limits. Taking first the momentum to zero and then the energy, corresponds to taking $h^*$ to $\pm \infty$ which gives zero contribution to the DSF. The opposite limit corresponds to $h^* \rightarrow 0$ and yields finite contribution  given by
\begin{align}
	\lim_{k\rightarrow 0} \lim_{\omega \rightarrow 0}S^{\twoph}(k,\omega) &= \frac{(2\pi)^2}{2}\fint_0 {\rm d}h_1 {\rm d}h_2 \,\vartheta_p(h_1) \vartheta_p(h_2) \rho_h(h_1) \rho_h(h_2)  \nonumber \\
	&\times\left( T^{\rm dr}(h_1, h_2)\right)^2 \frac{|v^{\rm eff}(h_1) - v^{\rm eff}(h_2)|^3}{\left(v^{\rm eff}(h_1)v^{\rm eff}(h_2)\right)^2}.
\end{align}
We note that the dependence on the order of the limits is caused by the form-factor itself and not by the density functions and therefore we expect the same feature also in more general, beyond the thermal equilibrium, settings.

\section{Application to correlation functions of higher densities and currents}~\label{sec:higher}

The computation presented in the previous section was performed for the specific density-density correlation function. However similar computations hold also for correlation functions of higher conserved densities and conserved currents given their similar analytic structure. Following~\cite{10.21468/SciPostPhys.6.4.049}, we recall that 
\begin{align}
	\langle \vartheta| \hat{q}_i |\vartheta, \bfp, \bfh\rangle = k(\bfp, \bfh) f_i(\bfp, \bfh), \\
	\langle \vartheta| \hat{j}_i |\vartheta, \bfp, \bfh\rangle = \epsilon(\bfp, \bfh) f_i(\bfp, \bfh),
\end{align}
where in the case of 2ph
\begin{align}
	f_i(p_1, p_2, h_1, h_2) &= \frac{2\pi \Tdr(h_2, h_1) h_i^{\rm dr}(h_2)}{k'(h_1) k'(h_2) (p_1 - h_1)} + \frac{2\pi \Tdr(h_1, h_2) h_i^{\rm dr}(h_1)}{k'(h_2) k'(h_1) (p_2 - h_2)} \nonumber \\
	&\,+ \frac{2\pi \Tdr(h_2, h_1) h_i^{\rm dr}(h_2)}{k'(h_1) k'(h_2) (p_2 - h_1)} + \frac{2\pi \Tdr(h_1, h_2) h_i^{\rm dr}(h_1)}{k'(h_2) k'(h_1) (p_1 - h_2)} + (\dots).
\end{align}

Following the procedure presented in the previous section, we start by computing the $(k,\omega)$-shell form-factors by setting $\bar{\bfp} = \bfh + \bar{\bfa}$ with $\bar{\bfa}$ given in~\eqref{alphabar}. Function $f_i(p_1, p_2, h_1, h_2)$ becomes
\begin{equation}
	f_i(p_1, p_2, h_1, h_2) = \frac{\Tdr(h_1, h_2)}{k} (v^{\rm eff}(h_1) - v^{\rm eff}(h_2))\left(\frac{h_i^{\rm dr}(h_1)} {\rho_t(h_1)  \Delta v_1^{\rm eff} } - \frac{h_i^{\rm dr}(h_2)}{ \rho_t(h_2) \Delta v_2^{\rm eff}}\right),
\end{equation}
where we introduced $\Delta v_i^{\rm eff} = v^{\rm eff}(h_i) - v^{\rm eff}(h^*)$. The $(k,\omega)$-shell form-factors are thus
\begin{align}
	\langle \vartheta| \hat{q}_i |\vartheta, \bfh, \bar{\bfp}\rangle &= \Tdr(h_1, h_2) (v^{\rm eff}(h_1) - v^{\rm eff}(h_2))\left(\frac{h_i^{\rm dr}(h_1)} {\rho_t(h_1)  \Delta v_1^{\rm eff} } - \frac{h_i^{\rm dr}(h_2)}{ \rho_t(h_2) \Delta v_2^{\rm eff}}\right) , \\
	\langle \vartheta| \hat{j}_i |\vartheta, \bfh, \bar{\bfp}\rangle &= v^{\rm eff}(h^*) \Tdr(h_1, h_2) (v^{\rm eff}(h_1) - v^{\rm eff}(h_2))\left(\frac{h_i^{\rm dr}(h_1)} {\rho_t(h_1)  \Delta v_1^{\rm eff} } - \frac{h_i^{\rm dr}(h_2)}{ \rho_t(h_2) \Delta v_2^{\rm eff}}\right).
\end{align}
These two classes of form-factors display different behaviour in the small $k$ and $\omega$ limits which translates into a different behaviour of the corresponding correlation functions. Recall, that taking first $k$ to zero and then $\omega$ to zero corresponds to taking $h^*$ to $\pm \infty$. In this limit, the charge density form-factor vanishes, whereas the current density form-factor approaches a finite value
\begin{equation}
	\lim_{h^* \rightarrow \pm \infty}\langle \vartheta| \hat{j}_i |\vartheta, \bfh, \bar{\bfp}\rangle = - \Tdr(h_1, h_2)\left(v^{\rm eff}(h_1) - v^{\rm eff}(h_2)\right)\left(\frac{h_i^{\rm dr}(h_1)} {\rho_t(h_1)} - \frac{h_i^{\rm dr}(h_2)}{ \rho_t(h_2)} \right).
\end{equation}
 In the opposite limit (first $\omega\rightarrow 0$ and then $k\rightarrow 0$), the form-factors have the opposite behaviour. That is the charge density form-factor takes a finite value
\begin{equation}
	\lim_{h^* \rightarrow 0} \langle \vartheta| \hat{q}_i |\vartheta, \bfh, \bar{\bfp}\rangle = \Tdr(h_1, h_2) (v^{\rm eff}(h_1) - v^{\rm eff}(h_2))\left(\frac{h_i^{\rm dr}(h_1)} {\rho_t(h_1)v^{\rm eff}(h_1)} - \frac{h_i^{\rm dr}(h_2) }{\rho_t(h_2) v^{\rm eff}(h_2)} \right),
\end{equation} 
whereas the conserved current form-factor vanishes. We will apply now these observations to the correlation functions.

Consider the connected two-point correlation functions of conserved charges and currents,
\begin{align}
	C_{ij}(x,t) = \langle \vartheta| \hat{q}_i(x,t) \hat{q}_j(0) |\vartheta\rangle, \quad \Gamma_{ij}(x,t) = \langle \vartheta| \hat{j}_i(x,t) \hat{j}_j(0,0) |\vartheta\rangle,
\end{align}
and their Fourier transforms $C_{ij}(k,\omega)$ and $\Gamma_{ij}(k,\omega)$. The $\twoph$ contributions to these correlation functions in the zero momentum and energy limit are
\begin{align}
	\lim_{\omega\rightarrow 0} \lim_{k\rightarrow 0}C_{ij}^{\rm 2ph}(k, \omega) &= 0, \\
	\lim_{k\rightarrow 0} \lim_{\omega\rightarrow 0}C_{ij}^{\rm 2ph}(k, \omega) &= \frac{1}{2}\fint_0 {\rm d}h_1 {\rm d}h_2 \,\vartheta(h_1) \vartheta(h_2) \rho_h(h_1) \rho_h(h_2) (\Tdr(h_1, h_2))^2 |v^{\rm eff}(h_1) - v^{\rm eff}(h_2)| \nonumber \\
	&\times \left(\frac{h_i^{\rm dr}(h_1)} {\rho_t(h_1)v^{\rm eff}(h_1)} - \frac{h_i^{\rm dr}(h_2) }{\rho_t(h_2)v^{\rm eff}(h_2)}\right) \left(\frac{h_j^{\rm dr}(h_1)} {\rho_t(h_1)v^{\rm eff}(h_1)}  - \frac{h_j^{\rm dr}(h_2) }{\rho_t(h_2)v^{\rm eff}(h_2)}\right),\\
	\lim_{\omega\rightarrow 0} \lim_{k\rightarrow 0}\Gamma_{ij}^{\rm 2ph}(k, \omega) &= \frac{1}{2}\int {\rm d}h_1 {\rm d}h_2 \,\vartheta(h_1) \vartheta(h_2) \rho_h(h_1) \rho_h(h_2) (\Tdr(h_1, h_2))^2 |v^{\rm eff}(h_1) - v^{\rm eff}(h_2)| \nonumber \\
	&\times \left(\frac{h_i^{\rm dr}(h_1)} {\rho_t(h_1)}  - \frac{h_i^{\rm dr}(h_2) }{\rho_t(h_2)}\right) \left(\frac{h_j^{\rm dr}(h_1)} {\rho_t(h_1)} - \frac{h_j^{\rm dr}(h_2) }{\rho_t(h_2)} \right),  \\ 
	\lim_{k\rightarrow 0} \lim_{\omega\rightarrow 0}\Gamma_{ij}^{\rm 2ph}(k, \omega) &= 0
\end{align}
To obtain these expressions we simply use our main result, eq.~\eqref{twoph_DSF}, and adopt the formula~\eqref{ko_ff} for $F_{\bfh}(k, \omega)$ to the form-factors of higher conserved densities and charges. Finally, we take the appropriate limit $h^* \rightarrow 0$ or $h^* \rightarrow \infty$.

We observe, as was already noted in~\cite{10.21468/SciPostPhys.6.4.049}, that the integrand in $\Gamma_{ij}^{\rm 2ph}(k, \omega)$ is a regular function (without the double poles) which turns the Hadamard integral into an ordinary one. This is not the case for the $C_{ij}^{\rm 2ph}(k, \omega)$ which still requires the regularization.

\section{The ground state DSF with 2ph}\label{sec:gs}

We specialize now back to the density-density correlation function and consider the $T=0$ limit of the 2ph contribution. In this case, the product of density factors $\vartheta(h_i) \rho_h(h_i + \bar{\alpha}_i)$, appearing in~\eqref{F_final} for $F_{\bfh}(k,\omega)$, localizes the integrals to the vicinity of the edge in the ground state distribution~\eqref{gs_vartheta},
\begin{equation}
	\vartheta(h_i) \rho_h(h_i + \bar{\alpha}_i) = \rho_t(q_F) |\bar{\alpha}_i| \left( \delta(h_i - q_F)\Theta(\alpha_i) + \delta(h_i + q_F) \Theta(-\alpha_i) \right)+ \dots,
\end{equation}
where $(\dots)$ are contributions of zero measure of the form $\Theta(x)(1 - \Theta(x))$. This leads to $2$ classes of contributions: with excitations on the opposite edges and with both excitations on the same edge. The form-factor $F_{\bfh}(k,\omega)$ vanishes when positions of holes coincide and therefore only the first class of excitations contribute. Given the symmetry in labelling the excitations we can assume that $h_1 = q_F$ and $h_2 = - q_F$. We will multiply the contribution to the correlation function by $2$ to account for the other choice. With this choice for the positions of the holes, we find
\begin{equation}
	\bar{\alpha}_1 = \frac{\omega + v_F k}{2 v_F k'(q_F)}, \qquad \bar{\alpha}_2 = - \frac{\omega - v_F k}{2 v_F k'(q_F)},
\end{equation}
where we used eq.~\eqref{alphabar} for $\bar{\alpha}_i$ and Fermi velocity $v_F \equiv  v^{\rm eff}(q_F)$.
Factor $\Theta(\alpha_1)\Theta(-\alpha_2)$ implements condition $\omega > v_F |k|$ with the form-factor equal zero otherwise. We find
\begin{align}
	G_{\bfh}(k, \omega) &= \rho_t^2(q_F) |\bar{\alpha}_1||\bar{\alpha}_2|\left(\frac{\Tdr(q_F, -q_F)}{2\pi}\right)^2 \times \delta(h_1 - q_F) \delta(h_2 + q_F) \Theta( \omega - v_F |k|).
\end{align}
This expression is valid when both $\bar{\alpha}_i$ are small compared to $q_F$. This implies that
\begin{equation}
	\omega + v_F |k| \ll 2 q_F v_F k'(q_F), \label{gs_condition}
\end{equation}
where we used that the non-zero contribution is possible only for positive energy $\omega$ (in fact the energy has to be at least $v_F |k|$). 
Recall that, according to~\eqref{F_final}, the full form-factor is 
\begin{equation}
	F_{\bfh}(k, \omega) = G_{\bfh}(k, \omega) \times \frac{k^4 |v^{\rm eff}(h_1) - v^{\rm eff}(h_2)|^3}{(\omega - v^{\rm eff}(h_1)k)^2(\omega - v^{\rm eff}(h_2)k)^2},
\end{equation}
which, including the mentioned above factor $2$, leads to
\begin{align}
	F_{\bfh}(k, \omega) &= \frac{1}{\pi^2} \left(\frac{\Tdr(q_F, -q_F)}{2\pi}\right)^2 \frac{v_F k^4}{|\omega^2 - v_F^2 k^2|} \times \delta(h_1 - q_F) \delta(h_2 + q_F) \Theta( \omega - v_F |k|), \label{2ph_GS}
\end{align}
where we used that $k'(h) = 2\pi \rho_t(h)$. To implement the Hadamard integral we need also the limiting expressions $F_{h^*, h_2}^{(1)}(k, \omega)$, etc. Setting $h_1 = h^*$ yields $\bar{\alpha}_2 = 0$ which leads to the factor $\vartheta(h_2) \rho_h(h_2)$ which is zero in the ground state. In turn $F_{h^*, h_2}^{(1)}(k, \omega) =0 $. Therefore, in the Hadamard integral, only the double integral term remains. The dependence on $h_1$ and $h_2$ is now only through the $\delta$-functions in~\eqref{2ph_GS}. The remaining integration is a product of 
\begin{equation}
	\lim_{\epsilon_1 \rightarrow 0} \int {\rm d}h_1 \delta(h_1 - q_F) \Theta(|h_1 - h^*| - \epsilon_1) = \lim_{\epsilon_1 \rightarrow 0} \Theta(|q_F - h^*| - \epsilon_1)  = \Theta( |q_F - h^*|),
\end{equation}
with an analogous expression for the second hole. The $\Theta$-functions encode the condition that there is no contribution when one of the excitations carries the whole energy and momentum which is equivalent to the condition already present in~\eqref{2ph_GS}. The $\twoph$ contribution to the DSF is then
\begin{equation}
	S^{\rm 2ph}(k, \omega) 
	= \frac{2 v_F k^4}{|\omega^2 - v_F^2 k^2|} \left(\frac{\Tdr(q_F, -q_F)}{2\pi}\right)^2 \Theta( \omega - v_F |k|).
\end{equation}
This expression is valid for $\omega$ and $k$ obeying~\eqref{gs_condition}.
The contribution exhibits a one-sided singularity of the form $(\omega - v_F |k|)^{-1}$ for $\omega$ approaching~$v_F |k|$. Presence of such edge singularity is expected from the general theory of non-linear Luttinger liquids~\cite{Imambekov_2008,Imambekov_2009,Imambekov_2009_1,Cherny_2009,2011_Shashi_PRB_84,Shashi_2012,2012_Imambekov_RMP_84}. Here, we see it in a simple setting as an effect of interactions between the linearly dispersing excitations at the opposite edges of the Fermi sea.

\section{Conclusions and outlook}

In this work, we have considered dynamic two-point correlation functions in the Lieb-Liniger model. For local operators conserving the number of particles, such correlation functions can be expanded in the thermodynamic form-factors and classified by a number of particle-hole excitations. Our focus was on analyzing the leading contribution from two particle-hole pairs, expressed by the Hadamard integral constrained by the energy and momentum conserving $\delta$-function. We have shown how to evaluate such integrals for the form-factors of the density operator $\hat{\rho}(x)$ and for higher conserved densities and currents present in integrable theories. In our analysis, we have focused on the leading contribution to those form-factors which come from small particle-hole excitations. Our results extend the findings of~\cite{10.21468/SciPostPhys.6.4.049} where the zero momentum and energy limit of the current-current correlation functions was derived. We have generalized them to finite, albeit small momenta and energies, and also considered two-point functions of conserved densities. In both cases, the resulting expressions involve double Hadamard integrals. We have also considered the zero-temperature limit of the dynamic density factor showing that it exhibits, as expected, edge singularities.  

In our considerations, we have focused only on the case of small momentum and energy with the diverging part of the $\twoph$ form-factor controlling the leading contribution. This is also the part of the form-factor that requires Hadamard regularization when performing the spectral sum. The subleading parts contain simple poles and regular terms. Hadamard integral over such terms reduces to the Cauchy principal value and regular integrals respectively. With this in mind, we can readily evaluate the $\twoph$ contribution of the full form-factor and not only its leading part. Given $\langle \vartheta| \mathcal{O}(0)|\vartheta, \bfp, \bfh\rangle$ with simple poles whenever $p_i \sim h_j$, we have that 
\begin{align}
	S^{\twoph}(k,\omega) &= \frac{(2\pi)^2}{4} \int {\rm d}\bfh \fint_{p_i = h_j} \!\!\!{\rm d}\bfp \,|\langle \vartheta| \mathcal{O}(0)|\vartheta, \bfp, \bfh\rangle|^2 \delta( \omega - \omega(\bfp, \bfh)) \delta( k - k(\bfp, \bfh)) \nonumber \\
	&= \frac{(2\pi)^2}{2} \fint_{h^*} {\rm d}\bfh\, \frac{\rho_p(h_1)\rho_p(h_2) \rho_h(h_1 + \alpha_1)\rho_h(h_2 + \alpha_2)}{k'(h_1)k'(h_2) |v^{\rm eff}(h_1) - v^{\rm eff}(h_2)|} |\langle \vartheta| \mathcal{O}(0)|\vartheta, \bar{\bfa} + \bfh, \bfh\rangle|^2,
\end{align}
where $\bar{\bfa}$ follows from solving the energy-momentum constraint
\begin{equation}
	k = k(\bar{\bfa} + \bfh, \bfh), \qquad \omega = \omega(\bar{\bfa} + \bfh, \bfh).
\end{equation}
This makes it possible to implement a numerical approach to the evaluation of such contributions when the full thermodynamic form-factor is known, e.g. for the density operator in the Lieb-Liniger~model. We also observe that this approach can be extended to higher particle-hole contributions given that the singular part of the higher form-factors follows the regular pattern. 

As this work was nearing completion, new results were reported based on the perturbative expansion of the density-density correlation function in $1/c$, appeared~\cite{2020arXiv200715396G}. The authors computed the correlation function on any finite entropy density state up to $1/c^2$ order. The $1/c^2$ terms involve contributions from $\twoph$ excitations which should enable the results of the two approaches to be compared in the future.  

\section*{Acknowledgments}

The author would like to acknowledge the support from the National Science Centre, Poland, under the SONATA grant~2018/31/D/ST3/03588.

\appendix

\section{1ph contribution} \label{app:1ph}

For the completeness of the presentation, we present in this Appendix the computation of the $\oneph$ contribution. For the case of the density-density correlation function these computations were performed in~\cite{SciPostPhys.1.2.015}. Here we present their straightforward generalization to a two-point function of any two conserved densities. The relevant form-factor is given by eq.~\eqref{oneph}. The form-factor has no singularities and the Hadamard regularization of the spectal sum is not needed. 
The $\oneph$ contribution to the correlation function of two conserved densities is
\begin{equation}
	C_{ij}^{\oneph}(k,\omega) =(2\pi)^2 \int {\rm d}p {\rm d}h \,\rho_h(p) \rho_p(h) h_i^{\rm dr}(h) h_j^{\rm dr}(h)  \delta(k - k(p,h))\delta(\omega - \omega(p,h)).
\end{equation}
We define $\alpha = p-h$ and rewrite the integration as
\begin{equation}
	C_{ij}^{\oneph}(k,\omega) =(2\pi)^2 \int {\rm d}h {\rm d}\alpha \rho_h(h + \alpha) \,\rho_p(h) h_i^{\rm dr}(h) h_j^{\rm dr}(h) \delta(k - k'(h) \alpha) \delta(\omega - v^{\rm eff}(h) k).
\end{equation}
Executing the integrals with the help of the $\delta$-functions yields,
\begin{equation}
	C_{ij}^{\oneph}(k,\omega) = 2 \pi \vartheta(h^*) \rho_h(h^* + \bar{\alpha}) \frac{ h_j^{\rm dr}(h^*) h_j^{\rm dr}(h^*)}{|k \kappa(h^*)|},
\end{equation}
where $h^*$ is a unique solution to $\omega = v^{\rm eff}(h) k$, $\bar{\alpha} = k/k'(h^*)$ and  $\kappa(h) = \partial_h v^{\rm eff}(h)$. We have also used that $k'(h) = 2\pi \rho_t(h)$. For a smooth distribution of rapidities, the shift $\bar{\alpha}$ can be neglected, which results in
\begin{equation}
	C_{ij}^{\oneph}(k,\omega) = 2\pi \vartheta(h^*) \rho_h(h^*) \frac{h_i^{\rm dr}(h^*) h_j^{\rm dr}(h^*)}{|k \kappa(h^*)|},
\end{equation}
with the dependence on $k$ and $\omega$ given implicitly by $h^*$ such that $\omega = v^{\rm eff}(h^*) k$. This representation makes it easy to compute the real space-time correlation function (strictly speaking the $\oneph$ contribution to it) with the result
\begin{equation}
	C_{ij}^{\oneph}(x,t)  = \vartheta(\tilde{h}) \rho_h(\tilde{h}) \frac{h_i^{\rm dr}(\tilde{h}) h_j^{\rm dr}(\tilde{h})}{|t \kappa(\tilde{h})|},
\end{equation}
with $\tilde{h}$ the unique solution to $x = v^{\rm eff}(\tilde{h}) t$.

In the ground state, the rapidities distribution have a sharp edge at Fermi rapidity $q_F$,
\begin{align}
	\rho_p(\lambda) &= \rho_t(\lambda) \left( 1 - \Theta(|\lambda| - q_F)\right), \nonumber \\
	\rho_h(\lambda) &= \rho_t (\lambda) \Theta(|\lambda| - q_F), 
\end{align}
Therefore the product $\vartheta(h^*) \rho_h(h^*) = 0$ and the non-zero contribution comes from the shift $\bar{\alpha}$. 
Consider first the case of positive momentum $k>0$. The particle-hole excitation must be then localized at the right Fermi edge: $h^* \leq q_F$ and $h^* + \bar{\alpha} \geq q_F$. We have then
\begin{equation}
	\vartheta(h^*) \rho_h(h^* + \bar{\alpha}) = \bar{\alpha} \rho_t(q_F) \delta(h^* - q_F) + (\dots) = \frac{k}{2\pi} \delta(h^* - q_F) + (\dots),
\end{equation}
where $(\dots)$ stands for terms of measure zero and terms of higher order in $\bar{\alpha}$. This yields the following contribution to the correlation function
\begin{equation}
	\frac{ h_i^{\rm dr}(q_F) h_j^{\rm dr}(q_F)}{|\kappa(h^*)|} \delta(h^* - q_F) = |k| h_i^{\rm dr}(q_F) h_j^{\rm dr}(q_F) \delta(\omega - v_F k),
\end{equation}
where we used that
\begin{equation}
	\frac{1}{|\kappa(q_F)|} \delta(h^* - q_F) = \delta(v^{\rm eff}(h^*) - v_F) = |k| \delta( \omega - v_F k),
\end{equation}
where $v_F = v^{\rm eff}(q_F)$ is the Fermi velocity.

In a similar fashion, for $k<0$, the excitation is localized at the left Fermi edge: $h^* \geq - q_F$ and $h^* + \alpha \leq -q_F$ and then
\begin{equation}
	\rho_p(h^*) \rho_h(h^* + \bar{\alpha}) = -\bar{\alpha} \rho_t^2(q_F) \delta(h^* + q_F) + (\dots) = \frac{|k|}{2\pi} \delta(h^* + q_F),
\end{equation}
with the contribution to the correlation function
\begin{equation}
	|k| h_i^{\rm dr}(q_F) h_j^{\rm dr}(q_F) \delta(\omega + v_F k).
\end{equation}
Adding the two contributions, we find
\begin{equation}
	C_{ij}^{\oneph}(k,\omega) = |k| h_i^{\rm dr}(q_F) h_j^{\rm dr}(q_F) \delta(\omega - v_F |k|).
\end{equation}
Finally, specializing to the DSF, $h_i^{\rm dr}(\lambda) = 2\pi \rho_t(\lambda)$, and using $2\pi \rho_t^2(q_F) = (N/L) v_F^{-1}$~\cite{KorepinBOOK}, we find
\begin{equation}
	S^{\oneph}(k,\omega) = 2 \pi \frac{N}{L} \frac{|k|}{v_F}  \delta(\omega - v_F |k|).
\end{equation}
The static correlator
\begin{equation}
	S^{\oneph}(k) = \int \frac{{\rm d}\omega}{2\pi} S^{\oneph}(k, \omega) = \frac{N}{L} \frac{|k|}{v_F},
\end{equation}
in agreement with the results of~\cite{SciPostPhys.1.2.015}.

\section{From the TBP to form-factors in the Lieb-Liniger model} \label{app:3ph}
In this Appendix, we conjecture the singular structure of the density operator form-factor in the Lieb-Liniger model with higher number of particle-hole excitations from the Thermodynamic Bootstrap Program~\cite{Bootstrap_JHEP,Cortes_Cubero_2020}. We adopt a shorthand notation for the form-factors
\begin{equation}
	f_{\vartheta}^{\mathcal{O}}(\theta_1, \dots, \theta_n) = \langle \vartheta | \mathcal{O}(0) | \vartheta, \theta_1, \dots, \theta_n\rangle.
\end{equation}

The thermodynamic bootstrap program for Integrable Quantum Field Theories predicts the singular structure of the form-factors when rapidities of the particles and holes coincide. In the relativistic notation, with $\theta$ being the rapidity, a form-factor of a local operator $\mathcal{O}$ with $4$ particles has the following singular structure
\begin{equation}
	f_{\vartheta}^{\mathcal{O}}(\theta_1, \theta_2, \theta_3, \theta_4) = - 2\pi \sum_{\sigma \in P_4} S_{\sigma}(\theta_1, \dots, \theta_4) \frac{\theta_{\sigma_3} - \theta_{\sigma_4} - i \pi}{\theta_{\sigma_1} - \theta_{\sigma_2} - i \pi} T^{\rm dr}(\theta_{\sigma_2}, \theta_{\sigma_4}) f_{\vartheta}^{\mathcal{O}}(\theta_{\sigma_3} \theta_{\sigma_4}). 
\end{equation}
We recall, that shifting a rapidity by $i \pi$, in a relativistic theory, corresponds to transforming it into a hole. For the correspondence with the Lieb-Liniger model, we choose $\theta_1 = h_1 + i \pi$, $\theta_2 = p_1$, $\theta_3 = h_2 + i\pi$ and $\theta_4 = p_2 $. This reduces the diverging part from $24$ terms to $4$ terms. We can moreover assume that $p_i \sim h_i$ so there are only 2 diverging terms. With these choices only two permutations matter: $\sigma = (1234)$ and $\sigma = (3412) = (23)(12)(34)(23)$. The latter leads to the following combination of the $S$-matrices
\begin{align}
	S_{(3412)}(\theta_1, \dots, \theta_4) &= S(p_1 - h_2 - i\pi)S(p_1 - p_2)S(h_1 - h_2) S(h_1 + i\pi - p_2) \nonumber \\ 
	&= \frac{S(p_1 - p_2) S(h_1 - h_2)}{S(p_1 - h_2) S(h_1 - p_2)}.
\end{align}
This expression, in the leading order in $p_i \sim h_i$, is equal to $1$. Therefore 
\begin{equation}
	f_{\vartheta}^{\mathcal{O}}(p_1, h_1, p_2, h_2) = - 2\pi T^{\rm dr}(p_1, p_2) \left( \frac{h_2 - p_2}{h_1 - p_1}  f_{\vartheta}^{\mathcal{O}}(h_2, p_2) + \frac{h_1 - p_1}{h_2 - p_2}  f_{\vartheta}^{\mathcal{O}}(h_1, p_1) \right). \label{app_ff}
\end{equation}
To simplify the notation we assumed here that $T^{\rm dr}(p_1, p_2)$ is a symmetric function of its arguments. 

Formally, the expression~\eqref{app_ff} is valid in the relativistic theory and to make a connection with the Lieb-Liniger model one would need to consider an appropriate non-relativistic limit~\cite{2009_Kormos_PRA_81,Kormos_2010}. Instead, we will assume that in all the terms we can simply substitute the corresponding Lieb-Liniger quantities. We will see that this leads to the correct result for the $\twoph$ form-factor and conjecturally also for higher ones. The only caveat, is the difference in the normalization of the form factors which in the Lieb-Liniger model are additionally divided by $(2\pi \rho_t(\lambda))^{1/2}$ for each particle or hole rapidity $\lambda$. Therefore, in the limit $p_i \sim h_i$ each particle-hole pair leads to the additional factor $2\pi \rho_t(h_i) = k'(h_i)$. Denoting the Lieb-Liniger form-factors with a bar, we find
\begin{equation}
	\bar{f}_{\vartheta}^{\mathcal{O}}(p_1, h_1, p_2, h_2) = - 2\pi T^{\rm dr}(p_1, p_2) \left(\frac{h_2 - p_2}{k'(h_1)(h_1 - p_1)}  \bar{f}_{\vartheta}^{\mathcal{O}}(h_2, p_2) -\frac{h_1 - p_1}{k'(h_2)(h_2 - p_2)} \bar{f}_{\vartheta}^{\mathcal{O}}(h_1, p_1) \right). 
\end{equation}
Recall that for $p_i \sim h_i$,
\begin{equation}
	k = k'(h_1) (p_1 - h_1) + k'(h_2) (p_2 - h_2).
\end{equation}
The terms $h_i - p_i$ appearing in the numerators, in the leading order of the diverging part, can be replaced by $k/k'(h_i)$. This gives
\begin{equation}
	\bar{f}_{\vartheta}^{\mathcal{O}}(\bfp_2, \bfh_2) = 2\pi k(\bfp, \bfh) T^{\rm dr}(p_1, p_2) \left(\frac{ \bar{f}_{\vartheta}^{\mathcal{O}}(h_2, p_2)}{k'(h_2) k'(h_1)(h_1 - p_1)}  + \frac{\bar{f}_{\vartheta}^{\mathcal{O}}(h_1, p_1)}{k'(h_1) k'(h_2)(h_2 - p_2)}  \right). 
\end{equation}
Finally, specializing to the conserved charge density, in the limit of small excitation $\bar{f}_{\vartheta}^{q_j}(h, p) = h_j^{\rm dr}(h)$, we find the singular part of the form-factor reported in~\eqref{twoph} for $p_i \sim h_i$. The full expression~\eqref{twoph} can be recovered by separately symmetrizing the labelling of holes and particles.

For $3$ particle-hole form factors, the TBP predicts the following singular structure in the limit of small ${\rm ph}$ excitations~\cite{Cortes_Cubero_2020}
\begin{align}
	&f_{\vartheta}^{\mathcal{O}}(\theta_1 + i \pi + \kappa_1, \theta_1, \theta_2 + i \pi + \kappa_2, \theta_2, \theta_3 + i \pi + \kappa_3, \theta_3) =\nonumber \\
	&= (2\pi)^2\left(\frac{\kappa_3}{\kappa_1} + \frac{\kappa_3^2}{\kappa_1 \kappa_2} \right)  T^{\rm dr}(\theta_2, \theta_3) f_{\vartheta}^{\mathcal{O}}(\theta_3 + i \pi \kappa_3, \theta_3) + (\text{cycl. perm.}),
\end{align}
where $(\text{cycl. perm.})$ refers to independent cyclic permutations of indices $\{1,2,3\}$ for particles and holes.
Isolating the most diverging part and following the same steps as for the $\twoph$ form-factor we conjecture the following expression for the Lieb-Liniger model
\begin{align}
	\bar{f}_{\vartheta}^{\mathcal{O}}(\bfp_3, \bfh_3) = \frac{(2\pi k(\bfp, \bfh)^2}{k'(h_1) k'(h_2) k'(h_2)} \left(\frac{T^{\rm dr}(p_1, p_3) T^{\rm dr}(p_2, p_3) \bar{f}_{\vartheta}^{\mathcal{O}}(h_3, p_3)}{(p_1 - h_1)(p_2 - h_2)}  + (\text{cycl. perm.})  \right).
\end{align}
Specializing to the conserved charge density, in the limit of small excitation $\bar{f}_{\vartheta}^{q_j}(h, p) = h_j^{\rm dr}(h)$, we find the form-factor reported in~\eqref{threeph}.

\section{Hadamard representation of the spectral sum} \label{app:Hadamard}

In~\cite{10.21468/SciPostPhys.6.4.049} it was argued that the spectral sum over the $\twoph$ excitations can be brought to the Hadamard form by an appropriate redefinition of the thermodynamic form-factor. In this appendix, we show that such redefinition is not necessary and the Hadamard regularization appears automatically when considering the thermodynamic limit of the spectral sum. We study the thermodynamic limit of the spectral sum following the approach introduced in~\cite{1742-5468-2010-11-P11012}.

We consider the $\twoph$ contribution to the finite-size correlation function, in which we dropped the subleading contributions in the system size,
\begin{equation}
	S_L^{\twoph}(x,t) = \frac{1}{4} \frac{1}{L^4} \sum_{p_1, p_2, h_1, h_2} e^{i k(\bfp, \bfh) x - i \omega(\bfp, \bfh)t} |\langle\vartheta| \hat{\rho}(0)|\vartheta, \bfp, \bfh\rangle|^2.
\end{equation}
The summations extend over possible particle-hole excitations over a chosen discretization of the thermodynamic state $|\vartheta\rangle$ with $N$ particles. The thermodynamic limit of $S_L^{\twoph}(x,t)$ does not depend on this choice of discretization. To rewrite the sum over holes as an integration we introduce the counting function $Q_p(\lambda)$ defined as
\begin{equation}
	Q_p(\lambda) = L \int_{-\infty}^{\lambda} {\rm d}\lambda' \rho_p(\lambda'),
\end{equation}
with the defining property that $Q_p(\lambda_j) = L j$ for $\lambda_j$ the $j$-th rapidity in the chosen discretization of $|\vartheta\rangle$. With the help of the counting function we rewrite the sum over $h_1$ as
\begin{align}
	\frac{1}{L}\sum_{h_1} F(h_1) &= \frac{1}{L}\sum_{j=1}^N \oint_{\gamma_j} {\rm d}z \frac{F(z) Q_p'(z)}{e^{2\pi i Q'(z)} - 1} \nonumber \\ 
	&= \frac{1}{L}\oint_{\Gamma} \frac{F(z) Q_p'(z)}{e^{2\pi i Q_p(z)} - 1} - \frac{2\pi i}{L} \sum_{k \in \{1,2\} }\underset{{z\rightarrow p_k} }{\rm res} \left( \frac{F(z) Q_p'(z)}{e^{2\pi i Q_p(z)} - 1} \right), \label{app_hole_sum}
\end{align}
where we used a shorthand notation $F(h_1) \equiv e^{i k(\bfp, \bfh) x - i \omega(\bfp, \bfh)t} |\langle\vartheta| \hat{\rho}(0)|\vartheta, \bfp, \bfh\rangle|^2$. The positively oriented contours $\gamma_j$ encircle $\lambda_j$ for each $j$, while avoiding $p_j$. In the second step, we combined all contours $\gamma_j$ into a single contour $\Gamma$ going around the real axis. In the process, we crossed the annihilation poles at $z = p_j$ and thus had to subtract their contributions. We are now in the position to take the thermodynamic limit. To this end, we observe that $Q_p'(z) = L \rho_p(z)$ and we define 
\begin{equation}	
	g_L(z) = \left(e^{2\pi i Q_p(z)} - 1\right)^{-1}.
\end{equation}
From $Q_p(x + i \epsilon) \approx Q_p'(x) + i \epsilon L \rho_p(x) + \mathcal{O}(\epsilon^2)$ it follows that
\begin{equation}
	\lim_{\rm th} g_L(x + i \epsilon) = \begin{cases}
		-1,&  \mbox{if } \epsilon > 0, \\
		0,&  \mbox{if } \epsilon < 0,
	\end{cases}\label{gL_limit}
\end{equation}
and the first term in~\eqref{app_hole_sum} gives
\begin{equation}
	\lim_{\rm th} \left( \frac{1}{L}\oint_{\Gamma} F(z) Q_p'(z) g_L(z) \right) = \int_{\mathbb{R} + i \epsilon} {\rm d}z\, F(z) \rho_p(z).
\end{equation}
For the computation of the second term we first interchange the thermodynamic limit with taking the residue,
\begin{equation}
	\lim_{\rm th} \underset{{z\rightarrow p_k} }{\rm res} \left(F(z) \rho_p(z) g_L(z) \right)= \underset{{z\rightarrow p_k} }{\rm res} \left( F(z) \rho_p(z) \lim_{\rm th} g_L(z) \right).
\end{equation} 
To compute the residue we need to know the thermodynamic limit of $g_L(z)$ for real $z = p_k$, which eq.~\eqref{gL_limit} does not provide. We can however compute it by representing $g_L(z)$ through the contour integral,
\begin{equation}
	g_L(z) = \frac{1}{2\pi i} \oint_{\gamma} {\rm d}z' \frac{g_L(z')}{z'-z}.
\end{equation}
This amounts to assigning to $g_L(z)$ the average value of $g_L(z')$ for $z'$ in the neighbourhood of $z$. 
For $z$ in the vicinity of the real number $p_k$ the contour $\gamma$ has to encircle $p_k$ and therefore extends above and below the real line. For large $L$, only the part of the contour above the real line, denoted $\gamma_+$, contributes and we find
\begin{equation}
	\lim_{\rm th} g_L(z) = - \frac{1}{2\pi i } \int_{\gamma_+} \frac{{\rm d}z'}{z' - z} = - \frac{1}{2}, 
\end{equation}
Therefore,
\begin{equation}
	\lim_{\rm th} \left( \underset{{z\rightarrow p_k} }{\rm res} \left(F(z) \rho_p(z) g_L(z) \right)\right) = - \frac{1}{2} \underset{{z\rightarrow p_k} }{\rm res} \left(F(z) \rho_p(z)\right),
\end{equation}
and the thermodynamic limit of the sum~\eqref{app_hole_sum} is
\begin{equation}
	\lim_{\rm th}\left(\frac{1}{L}\sum_{h_1} F(h_1)\right) = \underset{p_1, p_2}{\fint} {\rm d}x\, F(x) \rho_p(x).
\end{equation}
In writing this expression, we used the Hadamard regularization of the integral along the real axis defined as~\cite{doi:10.1063/1.4943300} 
\begin{equation}
	\underset{p}{\fint} {\rm d}x\, f(x) = \lim_{\epsilon \rightarrow 0^+} \int_{\mathbb{R} + i \epsilon} {\rm d}z\, f(z)  + i \pi \underset{{z\rightarrow p} }{\rm res} \left(f(z)\right),
\end{equation}
for a function with the second order pole at $p$. In the main text we use an alternative expression
\begin{equation}
	\underset{p}{\fint} {\rm d}x\, f(x) = \lim_{\epsilon\rightarrow 0^+} \left( \int_{-\infty}^{\infty} {\rm d}x f(x) \Theta(|x| - \epsilon) - \frac{2}{\epsilon} \lim_{x\rightarrow p} (x-p)^2 f(x) \right).
\end{equation}
The proof of their equivalence can be found in~\cite{doi:10.1063/1.4943300}.

The thermodynamic limit of the sum over the second hole $h_2$ can be analyzed in the same way and also leads to the Hadamard integral. The resulting expression, as a function of $p_1$ and $p_2$ is regular and summations over the particles' positions turn into the standard integrals. This procedure is straightforward to generalise to any number of particle-hole excitations which leads to the formula~\eqref{spectral} in the main text.

\bibliographystyle{JHEP}
\bibliography{2phDSF}

\end{document}